\documentstyle[preprint]{aastex}
\def\like{{\cal L}}

\newbox\grsign \setbox\grsign=\hbox{$>$} \newdimen\grdimen \grdimen=\ht\grsign
\newbox\simlessbox \newbox\simgreatbox \newbox\simpropbox
\setbox\simgreatbox=\hbox{\raise.5ex\hbox{$>$}\llap
     {\lower.5ex\hbox{$\sim$}}}\ht1=\grdimen\dp1=0pt
\setbox\simlessbox=\hbox{\raise.5ex\hbox{$<$}\llap
     {\lower.5ex\hbox{$\sim$}}}\ht2=\grdimen\dp2=0pt
\def\simgreat{\mathrel{\copy\simgreatbox}}
\def\simless{\mathrel{\copy\simlessbox}}

\begin{document}

\title{Statistical Analysis of Spectral Line Candidates \\
in Gamma-Ray Burst GRB870303}

\author{P.~E.~Freeman$^{1,6}$, C.~Graziani$^{1}$, D.~Q.~Lamb$^{1}$, \\
T.~J.~Loredo$^{2}$, E.~E.~Fenimore$^3$, T.~Murakami$^4$,\\
A.~Yoshida$^5$}
\altaffiltext{1}{Dept. of Astronomy and Astrophysics, University
of Chicago, Chicago, IL 60637}
\altaffiltext{2}{Dept. of Astronomy, Space Sciences Building, Cornell University
, Ithaca, NY 14853}
\altaffiltext{3}{Mail Stop D436, Los Alamos National Laboratory, Los Alamos, NM 87545}
\altaffiltext{4}{Institute of Space and Astronautical Science, 1-1, Yoshinodai 3-chome, Sagamihara, Kanagawa 229, Japan}
\altaffiltext{5}{Institute of Physical and Chemical Research, 2-1, Hirosawa, Wako, Saitama 351-01, Japan}
\altaffiltext{6}{Presented as a thesis to the Department of Astronomy and Astrophysics, The University of Chicago, in partial fulfillment of the requirements for the Ph.D. degree.}

\begin{abstract}
The {\it Ginga} data for the gamma-ray burst GRB870303 exhibit
low-energy dips in two temporally distinct spectra, denoted S1 and S2.
S1, spanning 4 seconds, exhibits a single line candidate at $\approx$ 20 keV,
while S2, spanning 9 seconds, exhibits apparently harmonically spaced line
candidates at $\approx$ 20 and 40 keV.  The centers of the time
intervals corresponding to S1 and S2 are separated by 22.5 seconds.
We rigorously evaluate the statistical evidence for these lines,
using phenomenological continuum and line models which in their details
are independent of the distance scale to gamma-ray bursts.
We employ the methodologies based on both
frequentist and Bayesian statistical inference that we develop in
Freeman et al.~(1999b).  These methodologies utilize the
information present in the data to select the simplest model that
adequately describes the data from among a wide range of continuum and
continuum-plus-line(s) models.  This ensures that the chosen model does
not include free parameters that the data deem unnecessary and that
would act to reduce the frequentist significance and Bayesian odds of
the continuum-plus-line(s) model.  We calculate the significance of the
continuum-plus-line(s) models using the $\chi^2$ Maximum Likelihood
Ratio test.  We describe a parametrization of the exponentiated
Gaussian absorption line shape that makes the probability surface in
parameter space better-behaved, allowing us to estimate analytically
the Bayesian odds.  We find that the significance of the
continuum-plus-line model requested by the S1 data is $3.6 \times
10^{-5}$, with the odds favoring it being 114:1.  
The significance of the
continuum-plus-lines model requested by the S2 data is $1.7 \times
10^{-4}$, with the odds favoring it being 7:1.  We also apply our
methodology to the combined (S1+S2) data.  The significance of the
continuum-plus-lines model requested by the combined data is $4.2
\times 10^{-8}$, with the odds favoring it being 40,300:1.
\end{abstract}


\section{Introduction}

The cause of gamma-ray bursts (GRBs) remains a mystery,
a quarter century after the announcement of their discovery by
Klebesadel, Strong, \& Olson (1973).
The recent discovery of optical transients associated 
with GRBs (e.g.~van Paradijs et al.~1997 and references therein), and
the apparent determination of redshifts for five of them$-$GRB970508
(Metzger et al.~1997), GRB971214 (Kulkarni et al.~1998), 
GRB980613 (Djorgovski et al.~1999), GRB980703 (Djorgovski et al.~1998),
and GRB 990123 (Kelson et al.~1999)$-$have indicated that some
(if not all) GRBs occur at cosmological distances.
(While Ockham's Razor might lead one to conclude on the basis of the
available evidence that all bursts are cosmological, it is important
to remember that the GRB sky location data themselves
do not yet rule out a separate galactic GRB source population; see, e.g.,
Loredo \& Wasserman 1995, 1998a,b.)
While broad-band observations indicate that
relativistically expanding fireballs can explain the
spectral and temporal behavior of these cosmological
transients (Goodman 1986; M\'esz\'aros \& Rees 1997),
it is the study of low-energy ($\simless$ 100 keV)
spectral line candidates seen in the spectra of other GRBs
that can potentially provide the most powerful means both to determine how
cosmological and/or galactic
GRBs occur and to place constraints on their environments.

Mazets et al.~(1980, 1981)
were the first to report low-energy spectral line candidates.
They found single dips and troughs in the spectra of 19 bursts
detected by the Konus detectors on {\it Venera 11} and {\it 
Venera 12}.\footnote{An additional burst with an apparent trough was later determined to be a solar flare; see Atteia et al. 1987.} 
This corresponds to $\approx 15$\% of the bursts
detected by Konus.  The statistical significances of these features
have not been reported.
Hueter (1987) then reported single low-energy dips
with modest statistical significance ($\sim 10^{-3}$)
in spectra of two bursts out of 21 detected by the {\it HEAO-1} A4 detector.
These reports influenced the design of the Los Alamos/ISAS
Gamma-Ray Burst Detector (GBD; Murakami et al.~1989) on the
{\it Ginga} satellite.
To help analysts differentiate spectral lines from changes in continuum
shape, a proportional counter (PC)
covering the energy range $\approx 1.5 - 30$ keV was included as part of
the GBD, in addition to a scintillator
counter (SC) covering $\approx 15 - 400$ keV.
(For Konus and {\it HEAO-1} A4, $E_{\rm low} \simgreat$ 20 keV.)
The spectra of three bursts observed by the GBD$-$GRB870303 
(spectrum S2), GRB880205 (spectrum b), and GRB890929$-$were found
to exhibit
apparently harmonically spaced absorption-like
line candidates at $\approx$ 20 and 40 keV
(Murakami et al.~1988, hereafter M88; Fenimore et al.~1988; 
Yoshida et al.~1991).
This is out of 23 bursts examined overall.
Another spectrum from an earlier epoch of GRB870303, denoted S1,
was found to
exhibit a single absorption-like line candidate at $\approx$ 20 keV
(Graziani et al.~1992, 1993, hereafter G92 and G93 respectively).
Analyses of the GRB880205 and GRB890929 spectra established the
significance of the line candidates 
to be $\approx 9 \times 10^{-6}$ and $\approx 3 \times 10^{-3}$,
respectively
(Fenimore et al.; Wang et al.~1989; Yoshida et al.).
An analysis of GRB870303
established the significance of
the line candidates in the spectra S1 and S2 to be
$\approx 1.1 \times 10^{-6}$ and $\approx 2.1 \times 10^{-4}$, respectively
(G92; we correct the values they report, as they used an
incorrect number of degrees of freedom when calculating significances).

Since {\it Ginga}, no GRB detectors
possessing the low-energy sensitivity of the GBD have flown.
Of those that have flown, the ones which are in principle the
most capable of detecting line candidates are the
eight Spectroscopy Detectors (SDs) of the Burst and Transient Source Experiment
(BATSE), on the {\it Compton Gamma-Ray Observatory}.
The gain settings of the individual SDs differ; those with the highest gain
settings can, in principle, observe GRBs at energies $\simgreat$ 10 keV.
An electronic artifact discovered after launch
affects energy calibration such that spectra are distorted in the first
$\approx$ 10 channels above the low-energy cutoff
(the so-called ``SLED'' effect; see Band et al.~1992).
While this can possibly affect line detection,
studies using simulated {\it Ginga} line candidate spectra
indicated that the 
BATSE SDs were still capable of detecting low-energy spectral line candidates
(Band et al.~1995).
However, no line candidates were definitively detected
during initial visual searches of those BATSE SD spectra with
the largest signal-to-noise ratios
(Palmer et al.~1994, Band et al.~1996).
The criteria for detection included having the candidate appear in
the data from at least one SD with $F$-test significance $\leq$ 10$^{-4}$,
with the contemporaneous data collected in other SDs being consistent
with the continuum-plus-line(s) model.
An automated line candidate search algorithm designed by the BATSE SD team 
(Briggs et al.~1996) was then
applied to spectra in 117 bright bursts for which there is at least one
spectrum with signal-to-noise $>$ 5 at $\approx$ 40 keV (Briggs et al.~1998).
This automated search, which is considerably more sensitive than
a visual search, yielded 12 candidate spectral line candidates for 
which the change in $\chi^2$ between the continuum and continuum-plus-line fits 
is $> 20$ (significance $< 5 \times 10^{-5}$). 
All candidates are emission-like lines at $\approx$ 40 keV, with
one absorption-like line candidate at $\approx$ 60 keV.
While Briggs et al.~estimate 
the ensemble chance probability of the most-significant feature
as $\simless$ 10$^{-3}$,  
and state that few of these features, if any, result from statistical
fluctuations, these should not be considered definitive detections, as
the contemporaneous data from other SDs is still being examined
(Briggs et al.~1999).

In sum, the {\it Ginga} observations provide strong evidence for spectral lines
that has as yet neither been independently confirmed, nor refuted.
There is, however, a theoretical bias against the
existence of lines, reinforced by the strong evidence 
supporting a cosmological distance scale for GRBs.
This has developed because few cosmological burst models have attempted to
account for the existence of harmonically spaced lines
(see, e.g., Stanek, Paczy{\'n}ski \& Goodman 1993, and Ulmer \& Goodman 1995,
who attempt to account for lines by invoking gravitational femtolensing).
However, the simple lack of theoretical models does not, nor cannot,
rule out the possibility of spectral lines in cosmological burst spectra.
In the galactic GRB paradigm, harmonically spaced absorption-like
lines are relatively simple to explain,
using cyclotron resonant scattering in the strong magnetic
field ($B \sim$ 10$^{12}$ G) of a neutron star.
Quantization of an electron's energy perpendicular to the magnetic
field $B$ facilitates the formation of
harmonically spaced lines with a spacing ${\Delta}E \approx$ 11.6$B_{12}$ keV.  
(See, e.g, Fenimore et al.; Wang et al.; Alexander and M\'esz\'aros 1989;
Miller et al.~1991, 1992; Isenberg, Lamb, \& Wang 1998; 
and Freeman et al.~1999a, hereafter Paper II.)

In this paper, we present rigorous methods of statistical
inference that the reader may use to firmly establish the evidence 
for spectral lines in GRB spectra, using
simple phenomenological models that are {\it independent of the
underlying physics of, and distance scale to, GRBs}.
To illustrate these methods, we apply them to
the spectral line candidates exhibited by the S1 and S2 spectra of
GRB870303.  In a companion paper (Paper II) we physically
interpret these line candidates within the galactic GRB paradigm,
using the cyclotron resonant scattering line transfer
code originally developed by Wang, Wasserman, \& Salpeter (1988).

In {\S}2 we describe the {\it Ginga} GBD and its observation of GRB870303.
In {\S}3, we present a basic introduction to the 
statistical concepts that we use in this paper.
These concepts are
discussed in greater detail in Freeman et al.~(1999b), hereafter Paper III.
In that work, we present general, rigorous, methodologies that address
the problem of establishing the existence of a line in a spectrum, that are
based upon both the 
so-called ``frequentist," and Bayesian, paradigms of statistical inference.
We apply both frequentist and Bayesian
methodologies in this work to ensure robust conclusions.
In {\S}4 we describe the method by which we 
select the simplest continuum model that
fits to the data outside the line candidate(s)
(rather than, e.g., simply assuming a continuum spectral shape).
We consider a wide range of spectral models, which assures that our
conclusions are robust.  Continuum model selection 
for {\it Ginga} GBD data is complicated by the presence of spectral
rollover at energies $\simless$ 5 keV, which we do not wish to model,
and we show how we adapt our
method to determine which PC bins may be included in fits.
In {\S}5, we describe how we select the simplest continuum-plus-line(s)
model that adequately fits to the data.
We introduce a parametrization of
the exponentiated Gaussian line in terms of its equivalent width $W_{\rm E}$
and full width at half maximum $W_{\rm \frac{1}{2}}$, the use of which
results in a
more well-behaved likelihood surface.  The model and its parametrization
allows us to treat saturated lines and, in addition, to apply analytic Bayesian
inference to both saturated and unsaturated lines.
We demonstrate the importance of applying models with as few free parameters as
possible, by applying saturated lines (with two, rather than three,
free parameters),
and/or by harmonically linking parameters between two lines,
in fits to these moderate resolution data.
We compare the selected continuum-plus-line(s) model to the
selected continuum model
to evaluate the frequentist statistical significance, and the
Bayesian odds in favor, of the best-fit continuum-plus-line(s) models
for GRB870303 S1 and S2.
We also determine the frequentist confidence
and Bayesian credible regions for the parameters of
these best-fit continuum-plus-line(s) models.
In {\S}6 we discuss our results.

\section{Observation of GRB870303}

We first summarize the characteristics of the {\it Ginga} GBD;
the interested reader will find more details in Murakami et al.~(1989).
The passively shielded and non-collimated
GBD contained two co-aligned instruments for detecting GRB photons.
The Proportional Counter (PC), used to detect low-energy photons,
consisted of a 3-cm deep Xe-CO$_2$
gas reservoir, with geometric area $\approx$ 63 cm$^2$.
The Scintillation Counter (SC), used to detect higher-energy photons,
consisted of a 1-cm thick NaI crystal with geometric area
$\approx$ 60 cm$^2$, backed by a 7.6-cm diameter phototube.
The entrance window of the SC was covered by a 0.2-mm-thick aluminum sheet,
whereas the entrance window of the PC was covered with a 
63.5-micron-thick layer of beryllium, which has greater transparency than
an aluminum layer of similar thickness at low energies.
In both instruments, an incident photon triggers an electron pulse;
the intensity of the pulse (the pulse height) is then used to discern
the amount of energy deposited by the photon.  Because the photon may
not deposit all its energy in these detectors, the PC and SC record the
number of counts as a function of photon {\it energy loss},
in 16 and 32 semi-logarithmically spaced bins, respectively.
To avoid the effect of uncertain
discriminator settings, we do not consider the lowest and highest
energy-loss bins in each detector (Murakami, private communication).
Excluding these bins, the PC and SC cover 1.4-23.0 keV and 16.1-335
keV, respectively, for the gain setting at the time that GRB870303
occurred.  
At the line candidate energy of $\approx$ 20 keV, the energy
resolution of the PC and SC are $\approx$ 3.4 and 5 keV, respectively;
at 40 keV, the resolution of the SC is $\approx$ 8.4 keV.

The GBD detected GRB870303 at 16:23 UT on 3 March 1987.  Figure 1
shows burst-mode time history data for the PC and SC.
The GBD continuously recorded burst-mode data at 0.5-second 
intervals.  These data were not stored in memory until a burst was detected,
at which time the data from 16 seconds
prior to the burst trigger until 48 seconds after the burst trigger
were stored.
The peak count rate (determined within a 4 second interval) in
the SC is $\approx$ 379 cts s$^{-1}$.  The background rate in the SC is
$\approx$ 572 cts s$^{-1}$.  
In addition to burst-mode data, the GBD also continuously recorded the
gamma-ray background in real-time mode, which had a coarse time
resolution (usually 16 seconds).  These data
are used to estimate the background count rate during the burst, in each
energy-loss bin.
By analyzing 150 seconds of real-time data from before the burst, and
220 seconds of real-time data from after the burst, we determine
that the background amplitude is constant as a function of time throughout
the burst interval.

The background-subtracted spectral data for GRB870303 exhibit line
candidates during two time intervals, the spectra of which we denote
S1 and S2 (following G92).  Figure 2
shows both spectra.  S1 is constructed from 4 seconds of data, during which the
burst had energy fluence
1.3 $\times$ 10$^{-6}$ erg cm$^{-2}$ in the bandpass 50-300 keV.
(This fluence is estimated from the best-fit model; Table 6.)
It exhibits a saturated line candidate at $\approx$ 20 keV.  
S2 is constructed from 9 seconds of data, during which the burst had energy
fluence 4.5 $\times$ 10$^{-6}$ erg cm$^{-2}$.  It exhibits two
harmonically spaced line candidates, at $\approx$ 20 and 40 keV.  
The midpoints of the time intervals from
which S1 and S2 are constructed lie 22.5 seconds apart.

Neither the PC nor SC could intrinsically determine
the angle of incidence of burst photons relative to the detector normal,
$\theta_{\rm inc}$.
The burst detector on the {\it Pioneer Venus Orbiter (PVO)} also observed
GRB870303; combining the photon time-of-arrival information from the
{\it Ginga} and {\it PVO} spacecraft limits the possible directions of the
burst to an annulus on the sky.  The burst photon 
angle of incidence is thus constrained
to lie within the range 
11.2$^{\circ}$ $\simless \theta_{\rm inc} \simless$ 57.6$^{\circ}$
(Yoshida, private communication, correcting Yoshida et al.~1989).  
In their analyses of the GRB870303 data,
M88, G92, and G93 assume $\theta_{\rm inc} =$ 37.7$^{\circ}$.
Since the shape and amplitude of a model counts spectrum that is
derived from a given photon spectrum depends sensitively on $\theta_{\rm inc}$,
we treat this angle as a freely varying model parameter in this work.
Because {\it Ginga} response matrices are computed using
computationally intensive Monte Carlo simulations, we use a grid
of fixed values 0.54 $\leq \cos{\theta_{\rm inc}} \leq$ 0.98,
with ${\Delta}(\cos{\theta_{\rm inc}})$ = 0.02.  This grid is
sufficiently dense to allow us to accurately determine 
statistical quantities such as line significance (see, e.g., Figure 6).

\section{Statistical Principles}

We analyze the line candidates in the spectra of GRB870303 S1 and S2
using both frequentist and Bayesian methods of model comparison and
parameter estimation.  In this section, we provide a basic introduction 
to those elements of frequentist and Bayesian
statistical inference relevant for the analysis of gamma-ray burst
spectral lines.  The reader will find more detail on these methods
in Paper III and references therein.

\subsection{Model Comparison}

\subsubsection{Frequentist Method}

The frequentist comparison of two models, the null hypothesis $H_0$ and the
alternative hypothesis $H_1$, is carried
out by constructing a test statistic $T$, which is usually a function
of the goodness-of-fit statistics for both models.
There are two probability distribution functions, or PDFs, which indicate
the a priori probability that we would observe the value $T$,
computed assuming the truth of $H_0$ and $H_1$, respectively.  The
test significance, $\alpha$, or Type I error,
is calculated by computing the tail
integral of the $H_0$ PDF from $T$ to infinity. 
The resulting number represents
the probability of selecting the alternative hypothesis $H_1$ when
in fact the null hypothesis $H_0$ is correct;
if the number is sufficiently
small, we reject $H_0$ in favor of $H_1$.  A common threshold
for rejecting the null hypothesis is $\alpha \leq 0.05$, though in
this work we use more conservative threshold values.

For the particular case of GRB spectral analysis,
the appropriate sampling distribution for the data is
the Poisson distribution, and the likelihood function ${\like}$, the
product of Poisson probabilities for the data in each bin, given model 
count rates, provides the best means to assess the viability of a model.
$H_0$ is the model with no line(s), $H_1$ is the model with line(s), and
$T = \frac{{\like}_{\rm max}(H_1)}{{\like}_{\rm max}(H_0)}$.
(The best-fit point, or mode, in parameter space is where the likelihood
function is maximized.)
To determine the $H_0$ PDF, one would simulate large numbers datasets from the
best-fit model for $H_0$ (i.e.~with the model parameters set to
best-fit values), and determine the distribution of observed
values of $T_{\rm sim}$.  After the $H_0$ PDF
is determined, finding the significance $\alpha$ is trivial.

However, this process may be computationally intensive. 
So the frequentist often
falls back upon the understanding that in the limit of a large number of
counts $n$ in a bin, the Poisson distribution is very nearly Gaussian with
a standard deviation ``root-$n$''.  
This understanding, in principle, allows
the use of Pearson's $\chi^2$ statistic, an approximation of $L = \log{\like}$,
to assess models:
\begin{equation}
s^2~=~\sum_{i=1}^{N} \frac{(m_i - n_i)^2}{\sigma_i^2} .
\end{equation}
The sum extends over $N$ data bins, and $m_i$ and $n_i$ are the
predicted and observed counts in bin $i$, respectively.  
The best-fit parameters for a given model are those for which 
$s^2$ is minimized.
This statistic has the advantage that analytic formulae may be
available to determine line candidate significance.
Under the same
assumption of a paraboloidal log-likelihood function in parameter space,
$s^2$ is sampled from the $\chi^2$ PDF
(in this paper, we follow the notation of Lampton, Margon, \& Bowyer 1976,
who reserve the symbol $\chi^2$ for a statistic which is explicitly
sampled from the $\chi^2$ distribution).  
There is some ambiguity in the choice of
$\sigma_i$: two widely used choices are $\sigma_i^2$ = $n_i$ (``data
variance''), and $m_i$ (``model variance'').  We denote fit statistics
using these two variance choices as $s_{\rm d}^2$ and $s_{\rm m}^2$,
respectively.

In this work,
we compare models $H_0$ and $H_1$ using the $\chi^2$ Maximum
Likelihood Ratio ($\chi^2$ MLR) test (Eadie et al.~1971, pp.~230-232).
In Paper III, we demonstrate that the use of this test results in
fewer Type I errors than both the $F$-test and the $\chi^2$
Goodness-of-Fit (GoF) test.  It is also the most powerful test
(Eadie et al., pp.~219-220).
In order to use it, the simpler model must be
nested within the more complicated alternative model, i.e.~the
simpler model must be obtainable by setting the extra ${\Delta}P\equiv
P_1-P_0$ parameters of the alternative model to
default values, often zero. The $\chi^2$ MLR test statistic is
${\Delta}s^2\equiv s^2(H_0)-s^2(H_1)$.  If the Gaussian
approximation is valid, $p({\Delta}s^2 \vert H_0)$  is given by the
$\chi^2$ distribution for ${\Delta}P$ degrees of freedom.

A sufficient condition that $s^2$ is distributed as $\chi^2$ 
($s^2 \sim \chi^2$) is that
$p(n_i \vert m_i)$ be Gaussian with mean $m_i$ and width $\sigma_i$. This
condition is not met if we fit a continuum-only model to data that has
a pronounced absorption-like or emission-like line candidate, 
regardless of the choice of variance:
we will be calculating the significance with
which $H_0$ is to be rejected in the regime where the $\chi^2$
approximation to the likelihood breaks down.
(This is because a second-order Taylor series expansion of the Poisson
log-likelihood, as a function of, e.g., $\delta \equiv 
\frac{\vert n_i - m_i \vert}{\sqrt{m_i}}$, is not sufficiently accurate when
$\delta \simgreat \sqrt{m_i}$; additional terms must be included.
It is precisely the second-order expansion which can be recast as
$\chi^2$.)  Consequently, the
significance calculated by looking up ${\Delta}s^2$ in the $\chi^2$
distribution with ${\Delta}P$ degrees of freedom cannot be expected to
agree with the ``true'' significance, i.e., the tail integral of the
true $H_0$ PDF determined using the Poisson likelihood function.
For the case relevant for this paper, absorption-like line candidates,
the use of model variances will cause the ``true" significance to be
underestimated,\footnote{To avoid semantical confusion: 
the smaller the value of $\alpha$, the greater the ``significance,"
in a qualitative sense.  Throughout this paper, we follow the convention
that $H_1$ becomes ``more significant" as $\alpha \rightarrow$ 0.}
while the use of variances derived from the data will lead to
overestimates of the ``true" significance
(i.e., $\alpha_{\rm \chi^2MLR,s_{\rm m}^2} > \alpha_{\like}$, and
$\alpha_{\rm \chi^2MLR,s_{\rm d}^2} < \alpha_{\like}$).
As shown in Paper III, we have found
that model variances provide a better estimate of the true significance,
so we calculate variances from the model in this work.

Another problem with the use of the $\chi^2$ MLR test is the
condition that estimates for the values of the additional
parameters introduced by $H_1$ must be drawn from normal distributions
(Eadie et al., p.~232), as
the line-centroid energy is drawn from uniform distribution
over the detector bandpass (e.g.~a spurious line could just as easily
be seen at 100 keV as 20 keV).
As discussed in Paper III, this tends to lessen the
significance of any detected line
(i.e.~$\alpha_{{\like},{\rm true}} > \alpha_{\like}$).
The magnitude of the decrease in significance is
dependent on the number of data bins and the width of the line. 
Our simulations indicate that for the specific case of the
{\it Ginga} GBD,
$\alpha_{\rm \chi^2MLR,s_{\rm m}^2} \approx \alpha_{{\like},{\rm true}}$.

\subsubsection{Bayesian Method}

As noted above, the appropriate sampling 
distribution for counts data is the Poisson distribution, and the likelihood
function ${\like}$, the product of Poisson probabilities for the data
in each bin, given model count rates, provides the best means to assess the
viability of a given model $M$.
The viability of a model in the frequentist method
is assessed in part by maximizing the likelihood function, but
in the Bayesian method, we integrate the likelihood function over 
the $P$-dimensional model parameter space.  The resulting quantity is called the
average likelihood:
\begin{equation}
p(D \vert M,I)~=~\int dx~p(x \vert M,I)~p(D \vert M,x,I)~=~\int dx~p(x \vert M,I)~{\like}(x) .
\label{eqn:num}
\end{equation}
In this equation, $D$ represents the data, while $x$ represents the freely
varying parameters of model $M$, and $I$ represents
information relevant to the analysis (e.g.~detector bandpass).
The likelihood is weighted at each point in parameter space by the 
conditional probability 
$p(x \vert M,I)$, called the prior probability, or simply
the prior.  The prior is
a quantitative statement of our state of knowledge about the
relative probability of each possible value of the parameter
$x$ before the data are examined.
There is a large body of literature on the subject of how to
assign priors (see Loredo 1992 and references therein), which we
will not summarize here.  When possible, we prefer to use
``least informative'' uniform priors (i.e.~constant amplitude functions), with
finite bounds which are determined in a physically meaningful way
(see Appendix B).

Bayes' Theorem allows us to calculate the posterior probability
$p(M \vert D,I)$ for
model $M$, given its average likelihood:
\begin{equation}
p(M \vert D,I)~=~p(M \vert I)\frac{p(D \vert M,I)}{p(D \vert I)}
\end{equation}
Here, $p(M \vert I)$ is the prior of the model itself (as opposed to the
values of each of its parameters), and $p(D \vert I)$ is a normalization 
factor.
A large posterior probability indicates support for the given model.
Instead of computing such a probability directly, we
determine the ratio of posterior probabilities for any two models within
a specified set of models $\{M_i\}$.  This quantity is called the {\it odds}:
\begin{equation}
O_{\rm 21}~=~\frac{p(M_{\rm 2} \vert D,I)}{p(M_{\rm 1} \vert D,I)}~=~\frac{p(M_{\rm 2} \vert I)}{p(M_{\rm 1} \vert I)} \frac{p(D \vert M_{\rm 2},I)}{p(D \vert M_{\rm 1},I)}~=~\frac{p(M_{\rm 2} \vert I)}{p(M_{\rm 1} \vert I)} B_{\rm 21} .
\end{equation}
Benefits of computing the odds are that we can ignore
the normalization $p(D \vert I)$,
and, if we do not have a priori preferences for either model,
the model priors $p(M_i \vert I)$.
The ratio of average likelihoods, denoted 
$B_{\rm 21}$ above, is termed the Bayes factor.
A Bayes factor of $>$10-20 is considered strong
evidence in favor of the alternative model, while a Bayes factor in
excess of 100 is considered decisive (see the review by
Kass \& Raftery 1995 and references therein).

Generally, the odds must be computed using numerical methods.  However,
if the shape of the likelihood surface in parameter space 
is similar to that of a 
multi-dimensional Gaussian function,
we may use the Laplace approximation to estimate 
the average likelihood (see, e.g., Kass \& Raftery,
Loredo \& Lamb 1992):
\begin{equation}
p(D \vert M_i,I)~=~p({\hat x_{i,{\rm 1}}} \cdot\cdot\cdot {\hat x_{i,P_i}} \vert I) (2\pi)^{P_i/2} \sqrt{\det V_i} {\like}_i^{\rm max}
\label{eqn:lap}
\end{equation}
$V_i$ is the covariance matrix,
determined by inverting the matrix
of likelihood function second derivatives evaluated at the mode.
To derive eq.~(\ref{eqn:lap}), we first assume that the prior does not
vary markedly around the mode,
so that the prior term in the integrand of eq.~(\ref{eqn:num}) 
may be replaced a constant evaluated at the mode.
(The hats placed on the parameters $x$ signify that we adopt
their values at the mode when evaluating the prior.)
What is left in the integrand
is the integral of ${\like}_i$, which we
assume has multi-dimensional Gaussian shape:
${\like}_{i,\rm max} \times G(x_i)$.  
Since $G(x_i)$ is an unnormalized Gaussian function, its
integral is $(2\pi)^{P_i/2} \sqrt{\det V_i}$.

We thus approximate the odds as
\begin{equation}
O_{21}~=~e^{{\Delta}L} (2\pi)^{{\Delta}P/2} \sqrt{ \frac{\det V_2}{\det V_1} } \frac{p({\hat {x}_{2,1}}\cdot\cdot\cdot{\hat {x}_{2,P_2}}\vert I)}{p({\hat {x}_{1,1}}\cdot\cdot\cdot{\hat {x}_{1,P_1}}\vert I)} .
\label{eqn:odds}
\end{equation}
(In this expression we use the log-likelihood, $L = \log{\like}$.)
This expression is sufficiently accurate-generally correct to within a factor
of two even if the likelihood surface
deviates somewhat from the ideal Gaussian shape-to allow us to
draw firm conclusions about the relative ability of the two models to
represent the observed data.

\subsection{Parameter Estimation}

\subsubsection{Frequentist Method}

We employ the method of projection to determine frequentist 
confidence intervals
(Eadie et al.; Lampton et al.).  For a given
parameter, we construct a set of values, and at each point on the set,
we minimize $s_{\rm m}^2$ with respect to the remaining parameters.  If the
shape of the likelihood surface in parameter space 
is similar to that of a multi-dimensional Gaussian function
(${\like} \propto \exp[-{\Delta}s_{\rm m}^2/2]$), and 
if the value of the likelihood at the mode is much larger than the 
maximum value of the likelihood along the parameter space boundary,
then the $n\sigma$ confidence interval for an
individual parameter is given by those values of the fixed parameter
such that $s_{\rm m,proj}^2$ = $s_{\rm m,min}^2$ + $n^2$.

If the likelihood surface deviates markedly from the ideal Gaussian shape,
such that it cannot be cast into that shape by parameter transformation,
or if the likelihood surface intersects
parameter space boundary near the mode,
the formula ${\Delta}s_{\rm m}^2 = n^2$ will
not apply.  (In the particular case of an otherwise
well-behaved likelihood surface which is
cut off at a parameter space boundary, the value ${\Delta}s^2$ which
defines the $n\sigma$ confidence interval is $< n^2$; the use
of ${\Delta}s_{\rm m}^2 = n^2$
will thus lead to overestimates of the confidence 
interval size.)  In these situations, we must perform simulations
to accurately determine the confidence intervals.  This is because the random 
variables in frequentist theory are the data, and not the model parameters:
we may not integrate over parameter space in an attempt
to make inferences about the parameters.

\subsubsection{Bayesian Method}

We may determine a Bayesian credible interval for a
particular parameter $x$ of model $M$, without reference to the other,
``uninteresting,'' parameters, collectively denoted $x'$, by
{\it marginalizing} the posterior function $p(x,x' \vert D,I)$ over the
space of parameters $x'$:
\begin{equation}
p(x \vert D,I) \propto \int dx' p(x,x' \vert I) p(D \vert x,x',I) .
\label{marg}
\end{equation}
We use the proportionality symbol because we ignore the normalization
factor $p(D \vert I)$.
The credible interval is defined as
\begin{equation}
z~=~\frac{\int_{x_1}^{x_2} dx p(x \vert D,I)}{\int_{{\rm all} x} dx p(x \vert D,I)} ,
\label{eqn:z}
\end{equation}
where $z$ is the desired probability content (e.g.~0.683 for 
1$\sigma$ bounds), and 
$p(x_1 \vert D,I) = p(x_2 \vert D,I)$.\footnote{
Here we implicitly assume the likelihood
surface is smooth and unimodal, so that any particular value
$p(x \vert D,I)$ occurs no more than twice.}  
Note that in the frequentist theory of confidence intervals,
$p(x_1 \vert D,I)$ does not have to equal $p(x_2 \vert D,I)$
(though they are equal in the projection method described above),
which means that confidence intervals, unlike credible intervals,
are not unique.
Only when the log-likelihood function is paraboloidal in parameter space
will the Bayesian method yield the same result as the frequentist 
projection method.

Even if the shape of the likelihood surface is nearly Gaussian,
numerical integration of the posterior to determine credible intervals
is preferable to the use of the Laplace approximation (eq.~\ref{eqn:odds}),
since the latter may not give sufficiently precise answers.
Numerical integration may not be feasible, however, if the
number of model parameters becomes too large ($\simgreat$ 5).
To use the Laplace approximation in this case, we would select
a grid of fixed values for the parameter $x$, and at each grid point
compute
\begin{equation}
p(x \vert D,I)~=~p(x,{\hat x}' \vert I)(2{\pi})^{P_i-1/2}\sqrt{{\det}V(x,{\hat x}')}{\exp[L(x,{\hat x}')]} .
\end{equation}
After we compute $p(x \vert D,I)$ at each point on the grid, we
would use eq.~(\ref{eqn:z}) as before to determine credible interval
bounds.

\section{Continuum Analysis}

In order to compare models with and without lines, and to estimate the
parameters of the lines (if lines are detected), we must specify a
continuum model.  The radiative processes that produce the continuum
spectrum of gamma-ray bursts are unknown.  Therefore, any physically
reasonable form for the continuum spectrum is a possibility.  And we
regard all models of burst continuum spectra, even when of a form (such
as power law or power law times exponential) produced by known
radiative processes, as purely phenomenological.  We consider a wide
range of possible continuum models, in order that we may draw
relatively robust conclusions from our study.
 
An unnecessarily complicated continuum model can reduce the frequentist
significance of a line, and the Bayesian odds favoring a model with a
line over one without.  {\it It is therefore important to select the
simplest continuum model that adequately describes the data.}  In this
section, we discuss the procedure that we use to select continuum models.

\subsection{Exclusion of the Energy-Loss Bins Affected by the Candidate Line}

In selecting the simplest continuum model that adequately describes the
data, we exclude those energy-loss bins associated with the line
candidate(s) from the fits, in order not to bias the outcome.
To determine which energy-loss bins to exclude from fits, we
first examine the raw data by eye to determine the approximate
line-centroid energy, $E_c$, of a candidate line.
An incident photon at this energy has probability
$p_i$ of being recorded as a count in the $i^{\rm th}$ energy-loss
bin.  If $p_i > 0.1$, we exclude the $i^{\rm th}$ bin from fits.
Using this
criterion, we exclude from the continuum model fits bins PC 14-15 and
SC 2-6 from the spectrum S1 and bins PC 14-15, SC 2-6, and SC 10-14 from the
spectrum S2.

\subsection{Continuum Model Selection}

In this section, we illustrate how we apply
frequentist statistical methodology to the
selection of best-fit continuum models for the S1, S2, and combined
(S1+S2) data.
Later, in our Bayesian analyses of the line candidates
exhibited by these data, we adopt the
continuum model selected using this frequentist method.
We do this because
the calculation of Bayesian odds favoring one
continuum model over another requires the stipulation of limits on the
allowed range of each continuum parameter, so that we may compute its prior.
Because the GRB continuum models are entirely phenomenological, it
is difficult to place meaningful, physically-motivated,
limits on the priors (e.g., how should we determine
the limits on a power law slope?).
We stress that the decision not to apply Bayesian methodology to
continuum model selection reflects our bias
against using subjectively chosen priors for model parameters, and should
not be viewed by the reader as an absolute injunction against
the use of the Bayesian methodology to select continuum models
when analyzing gamma-ray burst data.

We show our continuum model selection algorithm in Figure 3.
The selection of the best-fit continuum model is straightforward (Figure 3a).  
We fit each of a specified set of 
continuum models $M_i$ to the data.
(Because we exclude from these fits the
energy-loss bins in the vicinity of the line candidate(s), where the
model may not represent the data well, and thus where the approximations used
to derive $s_{\rm m}^2$ from ${\like}$ may be violated,
we can use the fitting statistic $s_{\rm m}^2$ [see {\S}3].)
If two or more models
have the same number of free parameters, we choose the one which fits
to the data with the lowest value $s_{\rm m}^2$.
Beginning with the simplest model (i.e.~the one with the fewest
number of free parameters),
we compute the significance of ${\Delta}s_{\rm m}^2$ for each alternative
model by computing 
$\alpha_{\rm {\chi^2}MLR}({\Delta}s_{\rm m}^2,{\Delta}P)$,
where ${\Delta}P$ is the number of additional free parameters introduced by
the alternative model.
If $\alpha_{\rm {\chi^2}MLR}$ is never $\leq$ 0.01,
we select the simplest model;
otherwise, we choose the simplest alternative model, and
compare that model against all remaining more complex alternative models.
We repeat this process until a continuum model is selected.

Complicating the process of continuum selection is the fact that
the magnitude of photon absorption in the beryllium window of
the GBD PC, at low energies ($E \simless$ 5 keV),
depends sensitively upon the burst photon incidence angle $\theta_{\rm inc}$.
As previously stated,
for GRB870303 this angle lies within the interval
$11.2^\circ \simless \theta_{\rm inc} \simless 57.6^\circ$.
Thus, for $E \simless$ 5 keV,
we cannot disentangle the absorption caused by the window from
any rollover intrinsic to the burst spectrum
and any absorption that may occur in intervening cold interstellar gas.
Modeling the spectral rollover thus greatly
complicates the fitting process while leaving our conclusions about the
line candidates essentially unaffected.
Hence, we add a step to the continuum selection algorithm which allows us
to determine which
energy-loss bins are most effected by the rollover (Figure 3b),
and we exclude these bins from subsequent line candidate analysis.
We model the spectral rollover using phenomenological absorption by a cold 
interstellar gas with a column density $N_H$ (model NH in Table 1).
We start by fitting to the data in all available PC bins (PC 2-13).
If the selected best-fit model includes the rollover parameter, we eliminate
the lowest-energy bin and repeat the process of model selection,
continuing until the data select a continuum model without rollover.

In our fits,
we consider four phenomenological continuum models (Table 1).
The four-parameter ``Band et al.~model'' (Band et al.~1993)
adequately describes all BATSE SD spectra to which it
has been applied.  The bandpasses of the BATSE SDs extend to much
higher energies than did the bandpass of the GBD SC, so the SC data
for GRB870303 may be insufficiently informative to require that the
the exponential cutoff energy, and/or second power law slope, of
this model be specified.
Thus, we consider two simpler models nested within the Band et al.~model:
a three-parameter power-law times exponential (PLE) model;
and a two-parameter power law (PL) model.
We also consider a two-segment broken power law (BPL) because
G92 and G93 use it to model the continuum of S2.  (This complicates
the model comparison process because the PLE model is not nested within
the BPL model, so that they are not directly comparable 
using the $\chi^2$ MLR
test.  But we never find the BPL and BPL+NH models to be the
best-fit models amongst models with 4 and 5 free parameters,
respectively.)
In our fits, we vary the logarithms of the normalization and energy
parameters, so that the shape of likelihood surface is more nearly Gaussian.
We also apply a ``pivot" energy of 20 keV (e.g.,
for the PL model, we use the formula
$\frac{dN}{dE}=A E^{-\alpha}=A' (E/20)^{-\alpha}$).
This helps reduce the size of the confidence and credible intervals for the 
highly correlated parameters $A$ and $\alpha$, while also improving
likelihood surface behavior.
Applying this set of models within our continuum selection algorithm,
we determine that
the PL and PLE models are the best-fit continuum models for S1
and S2 respectively (Table 2).

We use a similar algorithm to select continuum model(s) and bin
ranges for the fit to the combined (S1+S2) data.
This process of model selection is considerably more complicated
because we must both
determine whether the data explicitly request different values for some
continuum parameters common to both fits
(e.g.~power law slope, if we fit the PL model to S1 and the PLE model to
S2), and we also must determine
whether the data explicitly request separate burst photon incidence angles 
$\theta_{\rm inc}$ for the two datasets.
We determine that the data select the PL model for
S1 and the PLE model for S2, with separate normalizations and power law
slopes (Table 2).

We note that the PL model fits to the data of GRB870303 S1 with $s_{\rm m}^2$ 
= 18.22 for 28 degrees of freedom.  This value of $s_{\rm m}^2$ is
strikingly small: if $s_{\rm m}^2 \sim \chi^2$, then the probability of
finding this or a lower value of $s_{\rm m}^2$ is 0.080.
(G92,
who use $s_{\rm d}^2$, a different choice of energy-loss bins, and
assume $\theta_{\rm inc}$ = 37.7$^{\circ}$, compute a
probability 0.023.)
While this is not technically a significantly low value of $s_{\rm m}^2$,
we point out that extensive studies of the GBD were done which
demonstrate that instrumental
effects such as dead time, pulse pileup,
or bin overlap in the GBD did not conspire to lower the value of $s^2$
(e.g.~Graziani 1990).  Furthermore,
a detailed analysis of
GRB870725, a burst which occurred while {\it Ginga} was passing over the
Kagoshima Ground Station,
showed that the burst mode and real time data were still in complete
agreement nearly five months after GRB870303.

\section{Line Analysis}

\subsection{Line Model}

When fitting a candidate line in the spectrum of a GRB, one must choose
both a line model and a parametrization of that model.
Astrophysicists often use either an additive Gaussian line, $AL(E) =
C(E) - {\beta}G(E)$, or an exponentiated Gaussian line, $EL(E) =
C(E)\exp(-{\beta}G(E))$, to model the line.  
(We use the symbols $AL$ and $EL$ to denote line fluxes so as
to avoid confusion with the log-likelihood $L$.)
C(E) is the continuum flux, and
\begin{equation} 
G(E)~=~\exp\left( -{{(E-E_{\rm n})^2}\over{2\sigma^2}}\right) 
\label{eqn:gauss}
\end{equation} 
is the Gaussian line shape; $E_{\rm n}$ is the line-centroid energy of
the ${\rm n}^{th}$ harmonic line; and $\beta$ and $\sigma$ are the
unnormalized strength and width of the Gaussian. We use the
exponentiated line model rather than the additive Gaussian model because 
the flux in the latter can be negative, which is unphysical.

The exponentiated Gaussian model can be parametrized in different ways;
the choice of the parametrization affects the shape of the likelihood
surface in parameter space.  Figure 4
shows contours of
constant probability density as a function of ($\beta$,$\sigma$) from
fitting to an incident photon spectrum with a given set of line parameters.  
The contours show that in this particular case (and in general) the
likelihood surface will not have the shape of a
multi-dimensional Gaussian function; if it did have this shape, we
would observe elliptical contours.
(Note that the axes of the ellipses are not required
to be parallel to parameter axes.)
In frequentist statistics, the
parametrization of the line does not affect the calculated line
significance (which depends only on ${\Delta}s_{\rm m}^2$), but it does affect
the computation of confidence intervals.
The projection method gives a confidence region
that is accurate only if the shape of the likelihood surface closely
approximates that of a multi-dimensional Gaussian.
It is also advantageous because it greatly reduces the
computational burden of calculating credible regions using Bayesian
inference by allowing us to use the approximate expression in 
eq.~(\ref{eqn:odds}) to
determine the odds favoring the continuum-plus-line model.

While many parametrizations are well-behaved when the S/N of the
spectrum is large and/or the line is strong (but not saturated),
parametrizing the line in terms of its equivalent width, $W_{\rm E}$,
and full-width at half-maximum, $W_{\rm \frac{1}{2}}$, has many
advantages compared to parametrizing it in terms of $\beta$ and
$\sigma$.  First, we find that, when we take two-dimensional slices of
the parameter space and plot the probability density contours
corresponding to 2 and 3$\sigma$, parametrization of the line in terms
of $W_{\rm E}$ and $W_{\rm \frac{1}{2}}$ yields elliptical contours over a
much larger range of count rates than does parametrization in terms of
$\beta$ and $\sigma$ (see Figure 4).  Second, it has the added
advantage that it is more intuitive, in the sense that the visual shape
of the line is related more directly to $W_{\rm E}$ and $W_{\rm \frac{1}{2}}$
than to  $\beta$ and $\sigma$.  Third, parametrization of the line in
terms of $W_{\rm E}$ and $W_{\rm \frac{1}{2}}$ is useful because the $W_{\rm
\frac{1}{2}}$ of the line candidates in the spectra of gamma-ray bursts
is typically less than or of order the energy resolution of the
detector, so that the detector is sensitive to $W_{\rm E}$ but not to $W_{\rm
\frac{1}{2}}$ (see below).  We discuss the details of this
parametrization in Appendix A.

\subsection{Selection of the Line Model}

G93 point out that the standard line parametrization becomes
degenerate for saturated lines: for such lines, vast ranges of (very
large values of) $\beta$ and of (very small values of) $\sigma$ result
in lines which are virtually indistinguishable from each other using moderate
resolution NaI crystal spectrometers.   For example, if we convolve a
Gaussian line of width $\sigma$ with a Gaussian detector response of
width $\sigma_R$,  then for $\sigma \simless \sigma_{\rm R}$, the width
of the final Gaussian line is $\approx \sigma_{\rm R}$.  Thus a
saturated line may be adequately described by two parameters, its
line-centroid energy $E$ and its equivalent width $W_{\rm E}$.  How wide a
line must be before the third line parameter, the FWHM $W_{\rm
\frac{1}{2}}$, is requested by the data depends upon
how informative they are; it is more likely to be requested if the
S/N of the spectrum is high.  In the present context, this means that
the S2 data are more likely to request a third parameter than the S1 data.

This is important because inclusion of unnecessary line parameters
reduces ${\Delta}s_{\rm m}^2$ and ${\Delta}L$ per line parameter.  Since the
number of additional parameters in the continuum-plus-line(s) model
relative to the continuum model affects the frequentist significance,
and both the number of extra parameters introduced by the
continuum-plus-line(s) model and their prior ranges affects the
Bayesian odds, it is important to use the minimum number of line
parameters necessary to describe the data adequately.

The optimal search strategy for detecting a single narrow line is
therefore to fit to the data a two-parameter saturated line
parametrized by $(E,W_{\rm \frac{1}{2}})$, with the ratio $W_{\rm
E}/W_{\rm \frac{1}{2}}$ set to its maximum value, 1.015 (see Appendix
A and Figure 5). 
We then check whether or not the data is adequately described by a
saturated line by comparing this fit with one in which the line is
parametrized in terms of $E$, $W_{\rm E}$, and $W_{\rm \frac{1}{2}}$
(and $\beta$ is constrained to be $< \beta_o$).
To optimally detect apparently harmonically spaced lines, we want to reduce the
number of freely varying line parameters to the minimum requested by
the data.  
For two lines, the first step is to assume harmonic spacing between
the lines; this reduces the number of free parameters from six to five.
The next step is to assume that each line
is saturated; this reduces the number of free parameters from five to
three.  To reduce the
number of free parameters to two, we link the width of the first and
second harmonics.  For the purely historical reason that
line widths were once interpreted as Doppler widths of absorption 
profiles (e.g.~Fenimore et al.~1988),
we assume $W_{{\rm E},2}$ = 2$W_{{\rm E},1}$.  We could just as easily assume
$W_{{\rm E},2} = W_{{\rm E},1}$.  Since the width of the second
harmonic in S2 is smaller than the energy resolution at 40 keV,
the values of $s_{\rm m}^2$ and ${\like}$ 
are relatively insensitive to the assumed relation between
$W_{{\rm E},1}$ and $W_{{\rm E},2}$.

In Table 3 we list the line models that we consider when fitting
the spectra S1 and S2.

The procedure that we use to compare these continuum-plus-line models
is analogous to that used to compare continuum models, but with the
following  differences: we assume the continuum model and the range of
PC energy-loss bins that we selected using continuum model comparison;
and we restore to the fits the energy-loss bins that we excluded
earlier because they were near the energy of the line candidate.  We
note that not all the models we use to fit to S2 are nested within each
other, which precludes using the $\chi^2$ MLR test to
compute significances in some cases; however, the selection of the line
model was not effected by this.  Use of this procedure leads
to selection of the saturated line model (with parameters $E$ and $W_{\rm E}$)
for both S1 and S2.

Because $\cos{\theta}_{\rm inc}$ takes on discrete values in our analyses,
we must
apply a variation of eq.~(\ref{eqn:odds}) in our Bayesian method of
line model selection.  For any given value of 
$\cos{\theta_{\rm inc}}$, we use the Laplace approximation to estimate
the parameter space integral; we then sum these integrals over
all values of $\cos{\theta_{\rm inc}}$:
\begin{equation}
O_{21}~=~\frac{\sum_{\cos{\theta}_{\rm inc}} p({\hat x}_{2,1}{\cdot}{\cdot}{\cdot}{\hat x}_{2,P_2},\cos{\theta}_{\rm inc} \vert I)(2{\pi})^{P_2/2}\sqrt{{\det}V_2(\cos{\theta}_{\rm inc})}{\exp}(L_2^{\rm max}[\cos{\theta}_{\rm inc}])}{\sum_{\cos{\theta}_{\rm inc}} p({\hat x}_{1,1}{\cdot}{\cdot}{\cdot}{\hat x}_{1,P_2},\cos{\theta}_{\rm inc} \vert I)(2{\pi})^{P_1/2}\sqrt{{\det}V_1(\cos{\theta}_{\rm inc})}{\exp}(L_1^{\rm max}[\cos{\theta}_{\rm inc}])} .
\label{eqn:oddsreal}
\end{equation}
As noted in {\S}3, we assume uniform priors.  Appendix B describes
how we compute the prior for each model listed in Table 3, and Table 4
presents the formulae that we use to compute the prior for each model.
We need not specify priors for the continuum parameters; the use of the
exponentiated Gaussian line model allows us to factor the priors for
the line and continuum parameters, so that when we form the likelihood
ratio, the continuum priors cancel.
For the same reason, the prior for $\cos{\theta}_{\rm inc}$ cancels out
of the final expression.

The odds favoring Model S1-U over Model S1-S, and Model S2-B over
Model S2-A, is\footnote{In this paper, we follow the accepted Bayesian 
practice of treating the odds, the ratio of average likelihoods, as a singular
quantity.} $\simless$ 1:1, indicating a roughly 50\% chance that these
more complex models are the correct models to select.  An odds ratio of
1:1 falls far short of the 10:1 odds criterion
that would indicate sufficiently 
strong evidence in favor of the more complex models.

\subsection{Application to the Data of GRB870303}

\subsubsection{GRB870303 S1}

We estimate the frequentist significance of the spectral feature in GRB870303 S1
by comparing fits of the PL and PL+(S1-S) models to the data
(Tables 5-6, Figure 6).
The significance of the reduction in $s_{\rm m}^2$, 
for two additional parameters,
is $\alpha_{\rm \chi^2MLR}$ = 3.6 $\times$ 10$^{-5}$.
For reasons discussed above in {\S}3 and Paper III, this value is not
the ``true" significance that we would derive by simulating vast numbers
of datasets, but is expected to be approximately correct.

In our Bayesian analysis,
we apply the PL and PL+(S1-S) models to the data 
and use the modified Laplace approximation eq.~(\ref{eqn:oddsreal}) to
yield an estimate of the
odds favoring the continuum-plus-line model of 114:1 (Tables 5-6; 
Figures 6-7).
This is strong evidence in support of the line hypothesis.  
As discussed in {\S}3, the use of the Laplace approximation assumes that
a likelihood surface has ideal multi-dimensional Gaussian form, and
our reparametrization of the line model helps ensure that the likelihood
surface in this analysis has approximately
that ideal form.  The only way to ensure
accuracy of the Laplace approximation 
is to perform numerical integration as a check.  Unlike the
case for computing credible regions, where a portion of the numerical error
introduced by using sparse grids of parameter values will cancel out
because one is computing a ratio (eq.~\ref{eqn:z}; see Tierney \& Kadane 1986),
accurate numerical computation of the odds requires that we use 
a denser grid of parameter values.
Hence, we are computationally limited to performing 
numerical integration only over the five-dimensional parameter space of the
PL+(S1-S) model (the fifth parameter is the burst photon incidence angle).
This integration yields odds $\approx$ 130:1.  We conclude that the use of our
reparametrization and the Laplace approximation is accurate to well within
a factor of two.

In Table 7, we show the frequentist confidence and Bayesian 
credible intervals for the parameters of
the PL+(S1-S) model.  We find that for any given value $\cos\theta_{\rm inc}$
in the allowed range [0.54,0.98],
the confidence and credible intervals closely match, demonstrating the
efficacy of our line model reparametrization.  The likelihood
surface as a function of $\cos\theta_{\rm inc}$ is truncated (Figure 6);
while this does not affect the computation of Bayesian credible intervals,
it does cause the confidence intervals for 
$\cos\theta_{\rm inc}$ and any parameter correlated with
$\cos\theta_{\rm inc}$ (most notably the continuum normalization $A$
and slope $\alpha$) to be overestimated.

\subsubsection{GRB870303 S2}

We determine the frequentist 
significance of the spectral features in GRB870303 S2 
by comparing fits using the PLE and PLE+(S2-A) models to the data
(Tables 5-6; Figure 6).
The significance of the 
reduction in $s_{\rm m}^2$, for two additional parameters, is
$\alpha_{\rm \chi^2MLR}$ = 1.7 $\times$ 10$^{-4}$; the odds favoring
the continuum-plus-lines
model is 7:1 (Tables 5-6, Figures 6-7).
The line candidates in S2 are not
detected, if we apply the common criterion that the significance of the
candidate line must be $\leq$ 10$^{-4}$.  
The difference in odds between S1 and S2 is due to
the S2 data being more informative:
the errors on line parameters $E_{c,1}$ and $W_{{\rm E},1}$
are smaller for S2 than for S1, reducing the
average likelihood of the continuum-plus-lines model, and the odds.
In Table 8 we present frequentist confidence and Bayesian
credible intervals.  As seen in 
Figure 6, the likelihood surface as a function of 
$\cos\theta_{\rm inc}$ is truncated, leading to differences between the
computed intervals that arise for the same reason as stated above for S1.

\subsubsection{Joint Fits to the Combined (S1+S2) Data}

We fit to the combined (S1+S2) 
data because fits to this more informative dataset
can strengthen the statistical evidence favoring the line hypothesis.
We use the same continuum-plus-lines models that we applied to the S2 data
(Table 3), except that now for each model, we test
whether the data request different parameter values for
S1 and S2 (i.e.~we test Model S2-A with $E_{\rm 1,S1} = E_{\rm 1,S2}$,
then $E_{\rm 1,S1} \neq E_{\rm 1,S2}$, etc.).
Using both frequentist and Bayesian methods, we find
the best-fit joint line model is the S2-B model, with the values of
$E_{\rm 1}$ and $W_{\rm E,2}$ equal for S1 and S2, and
$W_{\rm E,1,S1} \neq W_{\rm E,1,S2}$.
(We note that the best-fit two parameter model, the S2-A model, was nested
within the best-fit three parameter model, the S2-B model, which in turn
was nested within the modified S2-B model shown above; hence the use of
the $\chi^2$ MLR test was valid at every step of frequentist model comparison.)

The frequentist significance of 
the reduction in $s_{\rm m}^2$, for four additional parameters, is
$\alpha_{\rm \chi^2MLR}$ = 4.2 $\times$ 10$^{-8}$, while the odds
favoring the continuum-plus-lines model is 40,300:1 (Tables 5-6; Figures 8-9).
In Table 9, we present frequentist confidence
intervals for the parameters of the continuum-plus-lines fit.
We find
that while the intervals are somewhat inaccurate because of
likelihood surface truncation, we can safely conclude that
the second harmonic line width is consistent with zero: there is
not overwhelming statistical evidence favoring the presence of the
second line.
We do not compute credible regions because the number of parameters
is too great.  We do not use the Laplace approximation to compute
approximate credible intervals because we sample the likelihood space
at only at discrete intervals in $\cos\theta_{\rm inc}$, which renders
difficult the computation of covariance matrices.

{\it We conclude that the joint (S1+S2) dataset thus
presents by far the strongest evidence supporting the 
hypothesis that spectral lines exist in gamma-ray burst spectra.}

\section{Discussion}

In this paper, we analyze the data of GRB870303 S1 and S2 using
rigorous statistical techniques developed for gamma-ray burst
line candidate analysis (Loredo \& Lamb 1992; G92; G93; 
Freeman et al.~1993, 1994; Paper III).  We conclude
that the line candidates exhibited by the S1 and S2 data have
significances 3.6 $\times$ 10$^{-5}$ and 1.7 $\times$ 10$^{-4}$, 
respectively,
with the Bayesian odds favoring the
continuum-plus-line(s) model being 114:1 and 7:1, respectively.  
Fits to the combined (S1+S2) data show
that the best-fit line model has significance
4.2 $\times$ 10$^{-8}$, with the odds favoring it being 40,300:1.
The results of these fits to the combined
data makes the line candidates they exhibit the most significant yet observed,
easily satisfying the most conservative line detection criteria.

The S1 and S2 data were previously analyzed by M88 and G92, 
who report significances for the line candidates in S2 of
$\sim$ 10$^{-3}$ and 2.1 $\times$ 10$^{-4}$, respectively. 
(We have corrected the significance computed by G92, because
they assumed the input number of degrees of freedom for the 
$\chi^2$~MLR model comparison test to be
the total number of parameters in the continuum-plus-lines model,
rather than the number of additional parameters used to parametrize
the lines.)
Both M88 and G92 assume $\theta_{\rm inc}$ = 37.7$^{\circ}$.
The line candidates in S2 are not
detected, if we apply the criterion that the significance of the
candidate line must be $\leq$ 10$^{-4}$ (see, e.g., Palmer 1994).  
G92 also discovered the line candidate in the spectrum
S1, computing its significance to be 1.1 $\times$ 10$^{-6}$ 
(also corrected).
G93 use a Bayesian method to analyze the line candidates,
and they report the odds in favor of the line
model to be 110:1 and 2.8:1 for S1 and S2 respectively.

The differences between the analyses of M88 and G92
and our analysis are summarized in Table 10.
Below, we discuss how each of these differences in turn
alters the computed significance and odds of the line
candidates.  Unless otherwise noted, we assume 
$\theta_{\rm inc}$ = 37.7$^{\circ}$ in all fits that we
perform below, to facilitate comparison between our results
and those derived previously.  We note that because of this assumption,
the derived significances, etc., stated below may differ
somewhat from analogous values presented in Table 5. 

\subsection{Choice of Frequentist Statistic}

We use the $s^2$ statistic with variances derived from the model count
rates in each energy-loss bin ($s_{\rm m}^2$).  
Both M88 and G92 use the $s^2$
statistic with variances derived from the data ($s_{\rm d}^2$). 
As noted in {\S}3 and demonstrated in Paper III,
the use of $s_{\rm d}^2$ can lead to an
overestimation of the significance of an absorption line, relative to
that derived using $s_{\rm m}^2$.  (Note that the exact opposite is true for
emission lines, where the use of $s_{\rm d}^2$ is to be preferred, if the
Poisson likelihood cannot be used.)
The magnitude of the difference between derived significances depends 
upon the size of the analyzed line candidate.
We find that if we fit to the S1, S2, and combined (S1+S2) data using
$s_{\rm d}^2$, the calculated line significances
are 3.3 $\times$ 10$^{-6}$, 1.5 $\times$
10$^{-5}$, and 1.5 $\times$ 10$^{-9}$, respectively.  If we
use $s_{\rm m}^2$,
the respective values are 1.2 $\times$ 10$^{-5}$,
4.0 $\times$ 10$^{-5}$, and 3.1 $\times$ 10$^{-8}$.

\subsection{Choice of Model Comparison Test}

In Paper III,
we use simulations to compute model comparison statistic PDFs 
for the $\chi^2$~MLR test, the $F$ test (used, e.g., by M88), and
the $\chi^2$ Goodness-of-Fit (GoF) test.  We use these PDFs to
determine that the $\chi^2$~MLR test is the most powerful test of
the three, i.e.~that the use of this particular test will result in
the highest rate of line detection, if
lines are present in the data.
Because it is the most powerful test,
we use the $\chi^2$~MLR test in this paper.

Application of the $F$ test (with test statistic
$\frac{P_2}{{\Delta}P}\frac{{\Delta}s_{\rm m}^2}{s_{\rm m}^2}$) 
to the S1, S2, and combined (S1+S2) data results in
significance estimates of
6.7 $\times$ 10$^{-6}$, 3.3 $\times$ 10$^{-4}$, and 
1.1 $\times$ 10$^{-7}$, respectively, as opposed to
1.2 $\times$ 10$^{-5}$, 4.0 $\times$ 10$^{-5}$, and 3.1 $\times$ 10$^{-8}$
for the $\chi^2$ MLR test.
The $F$ test
renders the candidate line in the S1 data more significant because of
the unusually small value of $s_{\rm m}^2$ for S1 (22.25 for 34 degrees of
freedom).
(We note that this does not make the $F$ test more powerful
in this particular case, because test power is computed from the PDF
of the model comparison test statistic calculated assuming
the truth of the alternative hypothesis, and not
from the results of fits to a single dataset.)

The application of the $\chi^2$ GoF test,
in which the $s_{\rm m,c}^2$ is compared to the $\chi^2$ distribution
for $N-P_c$ degrees of freedom,
to the S1, S2, and combined (S1+S2) data
leads to significances 0.15, 0.03, and 0.02.  The line candidates would not
be considered detected, if we assume a significance criterion 10$^{-4}$.
We note that if we apply the most generally-used 
threshold criterion of 0.05, the line candidate of
S1 would still not be detected.  This is a result of the unusually
small value of $s_{\rm m}^2$ for the continuum fit to S1.

\subsection{Use of Model Comparison to Select Continua}

In {\S}4, we describe the method with which we determine the 
best-fit continuum model for the S1, S2, and combined (S1+S2) data,
while also determining which
low-energy PC bins to include in fits.
Fenimore et al.~(1988), in their analysis of GRB880205,
were the first to apply a number of different
continuum models to data.  They adopt a three-segment power law as
representative of all possible models,
after determining that the choice of continuum has little effect upon
the detection of lines in these data.
An antecedent of the method we prescribe in this
paper was used by G92 and G93, but they did not test either the
PLE or Band et al.~models,
nor did they use model comparison to determine the usable range
of PC bins.
In both works, PC 10 is
adopted as the lowest usable bin ($E_{\rm low} \approx$ 5.7 keV).

In Table 11, we show how the results of fits to S1 and S2
change if we apply the BPL model used by G92, and the Band et al.~model,
to the data.  We also examine how our results change if we
limit ourselves to fitting to the data in PC 10-14 only, the PC bin
range used by G92.
We find that using the Band et al.~model leads to {\it increases}
in line significance, most notably for the fit to S1.
While initially encouraging, this result does {\it not} actually strengthen the
evidence supporting the existence of lines in these data, despite the
strong Bayesian prior supporting the Band et al.~model.
This is because in addition to the model prior, we must take into account
our prior expectation for the values of each model parameter, either
qualitatively or quantitatively.  In the particular case of the fit to the
data of S1, the inclusion of 
the (unrequested) second power-law segment causes
the best-fit model parameter values to deviate strongly
from their expected values.
For instance,
the best-fit exponential cutoff energy does not lie within its characteristic
range ($>$100 keV); instead, it is $\approx$ 10 keV.
The model is attempting to fit what remains
of the (no longer statistically significant)
low-energy spectral rollover at energies $\approx$ 5 keV (Figure 10).
This causes an increase in the continuum flux at 20 keV,
increasing both the equivalent width needed to fit the data in the line region
(from 9.84 to 10.7 keV)
and the significance of the line.
The increase in significance is greatly reduced when the lowest energy
bins PC 8-9 are removed from the fit.  Also, we find that we
cannot compute the odds
favoring the continuum-plus-line model if we use the Band et al.~model.
Parameter values along the Band et al.~model space boundary defined by
$\alpha_1 = \alpha_2$ are highly probable with respect to the mode
(because the slope of the unrequested second power-law segment
tends towards the slope of the first power-law segment), and because
of this boundary, our fitting program was unable to estimate the covariance
matrix values for the model parameters.

The fit of the Band et al.~model to the S1 data demonstrates how
including unjustified continuum parameters in the fit can alter
the computed line significance.
We feel that Briggs et al.~(1998) provide another demonstration
with their analysis of the emission-like line candidate of GRB941017,
which was observed by the BATSE SDs.
Their philosophy differs from that espoused in this paper:
they contend that to demonstrate the existence of a line in this (or
any) burst, one
must show that the data require the line regardless of whichever reasonable
continuum model is assumed.
The use of a Band continuum model plus two-parameter line to fit
the GRB941017 data collected by BATSE SD 0 results in a line
candidate significance of 7 $\times$ 10$^{-5}$.
After adding a low-energy spectral break to the continuum (which introduces
two additional continuum parameters), the significance is
decreased to 0.04.
Thus they feel that they cannot prove the spectral feature is a line.
We feel that there are two problems with this approach, one observational,
and the other methodological.
First, such breaks have not been observed in any other continuum
spectrum, in particular those of bursts observed by the {\it Ginga} GBD,
whose low-energy coverage is superior to that of the BATSE SDs.
Hence we feel that the proposed continuum shape may not be
reasonable.  Second, 
Briggs et al.~do not use model comparison to
justify that the additional continuum parameters introduced with the
low-energy power law segment are
necessary to adequately fit the data.  

\subsection{Use of Model Comparison to Select Line Models}

Fitting to data with models that have more free parameters than 
necessary can lead to a marked reduction in derived line significance.
The moderate resolution data collected by
the {\it Ginga} GBD lack the informative power to require that
each line candidate be fit with a line model parametrized by
line-centroid energy, equivalent width, and full-width at half-maximum,
with each value freely varying.
This fact has come to be recognized through successive analyses of
{\it Ginga} GBD data.
Initially, analyses of harmonically spaced line candidates
by M88, Fenimore et al., and G92\footnote{
Specifically, for their computation of the line candidate significance for
the S2 data.}
featured models containing
two lines with six independently-varying parameters.
Yoshida et al.~1991 and G92\footnote{For their
estimation of parameter values for S2.} reduce the number of
free parameters to five,
by testing models in which the line-centroid
energies are harmonically related ($E_2 = 2E_1$).
A further step towards model simplification is taken by 
G93, who test a four-parameter line 
model with harmonically related values of
$E_{\rm n}$ and $W_{\rm \frac{1}{2},n}$;
they also test a two-parameter line
model in fits to the S1 data, assuming that the line is saturated
($W_{\rm \frac{1}{2}} \approx W_{\rm E}$).
In this work, we push model simplification to its limits, by testing both
the saturated line model of G93 in fits to the S1 data and the 
two-free-parameter S2-A model in fits to S2.

In Table 12, we demonstrate the effect of including
unnecessary line model parameters in fits to the S1 and S2 data, fitting
the former with the three-parameter unsaturated line model (S1-U), and the
latter with
five- and six-parameter models.  For S1, the addition of a parameter causes
no change in $s_{\rm m}^2$ (because the mode lies at a parameter space
boundary), whereas for S2, we find
that each additional parameter lowers $s_{\rm m}^2$ by $\approx$ 1, which is
just what we expect if we include in the fit parameters that the data
do not request. 

\subsection{Effect of the Parametrization and Choice of Prior on Bayesian Odds}

Our Bayesian analysis differs from that reported in G93 in that
we scale the continuum amplitude and energy cutoff parameters
logarithmically.  This helps create a
likelihood surface that more closely resembles a
multi-dimensional Gaussian, and thus makes the estimation of the
covariance matrix more accurate.
We find that while this change has little effect
upon the odds favoring the line hypothesis for S1,
it does increase the
odds for S2 by nearly factor of two (from 8.7:1 to 14:1, for the S2-A model
with $\theta_{\rm inc}$ = 37.7$^{\circ}$, and for our chosen PC bin range).

\acknowledgements

The authors would like to thank the referee, David Band,
for his careful reading of the text and his many helpful comments.
This research was supported in part by NASA Graduate Student Research
Fellowships NGT 50778 (PEF) and 50617 (CG), and NASA Grants NAGW 830,
NAGW5-1868, and NAGW5-1464.

\appendix

\section{The $W_{\rm E}$-$W_{\rm \frac{1}{2}}$ Parametrization}

The equivalent width, $W_{\rm E}$, 
of an exponentiated Gaussian line $G(E)$ in eq.~(\ref{eqn:gauss}) is
\begin{eqnarray}
W_{\rm E}(E,\beta,\sigma)~&=&~\int_0^{\infty} dE \frac{C(E)-EL(E)}{C(E)} \nonumber \\
   &=&~\int_0^{\infty} dE (1~-~\exp[G(E)]) \label{eqn:a1} \\
   &=&~\sqrt{2} \sigma \Phi(E,\beta,\sigma) \nonumber,
\end{eqnarray}
where
\begin{equation}
\Phi(E,\beta,\sigma)~=~\int_{\frac{-E}{\sqrt{2}\sigma}}^{+\infty} dx \left[1 - \exp (-\beta e^{-x^2})\right]~\approx~\int_{-\infty}^{+\infty} dx \left[1 - \exp (-\beta e^{-x^2})\right] .
\label{eqn:a2}
\end{equation}

The approximation $\Phi(E,\beta,\sigma) \approx \Phi(\beta)$
is satisfied in all cases where we infer a low-energy
line candidate in data, breaking down only as $\beta \rightarrow 
\exp(\frac{E}{\sqrt{2}\sigma})$.  In the following, the limit 
$\beta \rightarrow \infty$ has the meaning of $0 \ll \beta < 
\exp(\frac{E}{\sqrt{2}\sigma})$.

As $\beta$ $\rightarrow$ 0, the function $\Phi(\beta)$ has the limit
\begin{eqnarray}
\lim_{\beta \rightarrow 0} \Phi(\beta)~&=&~\int dx \beta e^{-x^2} \nonumber \\
                                       &=&~\beta \sqrt{\pi} .
\label{eqn:a3}
\end{eqnarray}
As $\beta \rightarrow \infty$,
the approximate integrand in eq.~(\ref{eqn:a2}) behaves 
approximately as a box function $B(x)$, where
\begin{equation}
B(x)~=~\cases{ 1,&if $\vert x \vert$ $<$ $x_0$ ;\cr
           0,&otherwise.\cr} 
\end{equation}
We estimate $x_o$ by assuming that $\beta e^{-x_0^2}$ = 1, or
\begin{equation}
x_0~=~\sqrt{\log \beta} .
\end{equation}
Thus
\begin{equation}
\lim_{\beta \rightarrow \infty} \Phi(\beta) \approx 2\sqrt{\log \beta} .
\label{eqn:a6}
\end{equation}

Differentiating $\Phi(\beta)$, we find that for all $\beta$,
\begin{equation}
\frac{d\Phi}{d\beta} > 0 ~~~~~{\rm and}~~~~~ \frac{d^2\Phi}{d\beta^2} < 0 .
\end{equation}
We may invert the mapping $\beta
\rightarrow \Phi(\beta)$ for all values of $\beta$,
although for even moderate values of $\Phi$ the corresponding value of
$\beta$ may be very large.  It is numerically straightforward to
invert $\Phi$ and conclude the reparametrization 
($\sigma$,$\beta$) $\rightarrow$ ($\sigma$,$W_{\rm E}$).

We use the physical full-width at half-maximum of the line itself 
({\it not} of the Gaussian) to define
a more physically meaningful parametrization:
\begin{equation}
\frac{1 - \exp [G(E + W_{\rm \frac{1}{2}}/2)]}{1 - \exp[G(E)]}~=~\frac{1}{2} .
\end{equation}
We invert this equation, and use
eq.~(\ref{eqn:gauss}), to write $W_{\rm \frac{1}{2}}$ as:
\begin{equation}
W_{\rm \frac{1}{2}}~=~2\sqrt{2} \sigma \left\{ \log (\beta) - \log \left[\log \left(\frac{2}{1 + e^{-\beta}}\right)\right]\right\}^{ \frac{1}{2} } .
\end{equation}
It follows that
\begin{equation}
\lim_{\beta \rightarrow 0} W_{{{1}\over{2}}}~=~2 \sigma \sqrt{2\log 2} ,
\label{eqn:a10}
\end{equation}
and that
\begin{equation}
\lim_{\beta \rightarrow \infty} W_{{{1}\over{2}}}~=~2 \sigma \sqrt{2(\log \beta - \log (\log 2))} . 
\label{eqn:a11}
\end{equation}

The reparametrization ($\sigma$,$\beta$) $\rightarrow$
($W_{\rm E}$,$W_{{{1}\over{2}}}$) is simplified by the 
observation that the ratio $r$ = $W_{\rm E}~W_{{{1}\over{2}}}^{-1}$ depends only
on $\beta$.
From eqs.~(\ref{eqn:a1}), (\ref{eqn:a3}), and (\ref{eqn:a10}), we find
\begin{equation}
\lim_{\beta \rightarrow 0} {{W_{\rm E}}\over{W_{{{1}\over{2}}}}}~=~0 ,
\end{equation}
while from eqs.~(\ref{eqn:a1}), (\ref{eqn:a6}), and (\ref{eqn:a11}), we find
\begin{equation}
\lim_{\beta \rightarrow \infty} {{W_{\rm E}}\over{W_{{{1}\over{2}}}}}~=~1 .
\end{equation}

The ratio $W_{\rm E}/W_{\rm \frac{1}{2}}$ is nearly, but not quite, a
monotonic function of the unnormalized Gaussian amplitude $\beta$: it rises
sharply from zero, reaching $W_{\rm E}/W_{\rm \frac{1}{2}}$ = 1 when
$\beta \approx$ 4.75 and peaking at $\approx$ 1.015 when $\beta =
\beta_o \approx$ 18.7, 
before tapering off to 1 as $\beta \rightarrow \infty$
(Figure 5).  Thus the line
begins to saturate when $\beta \approx$ 4.75 and approaches a square well
shape as $\beta \rightarrow \infty$.
An observed line has a value of $W_{\rm E}/W_{\rm \frac{1}{2}}$ which falls
in the range 0 $\leq W_{\rm E}/W_{\rm \frac{1}{2}} \leq$ 1.015 (and therefore
$\beta$ in the range 0 $\leq \beta \leq \beta_o$) unless it is highly
saturated.  We constrain $\beta$ to this range, with little loss in generality,
as statistical fits made without this constraint will differ very little from
those made with it, unless the S/N of the line is extremely large.

\section{Bayesian Prior Probability Distribution for the $W_{\rm E}$-$W_{\rm \frac{1}{2}}$ Parametrization}

There are no standard rules for determining the range and shape of
the prior probability distribution in Bayesian methodology
(see, e.g., Loredo 1992 and references therein).
In this paper, we seek to assign priors that are ``least
informative.''  We assume a uniform, i.e.~flat, distributions,
bounded in a physically meaningful way.

For a single line, we use the product rules of 
probability theory to expand the prior:
\begin{equation}
p({\hat E},{\hat W_{\rm E}},{\hat W_{\rm \frac{1}{2}}} \vert I)~=~p({\hat W_{\rm \frac{1}{2}}} \vert {\hat W_{\rm E}},{\hat E},I)~p({\hat W_{\rm E}} \vert {\hat E},I)~p({\hat E} \vert I) .
\end{equation}
$I$ represents background information about the experiment, such as
detector bandpass, which allows us to
specify a flat and bounded distribution of the line-centroid energy $E$:
\begin{equation}
p({\hat E} \vert I) = \frac{1}{E_{\rm high} - E_{\rm low}} .
\label{eqn:b2}
\end{equation}
If ${\hat E} <$ $\frac{1}{2}E_{\rm high}$,
we may specify $p({\hat W_{\rm E}},{\hat E} \vert I)$ by assuming 
\begin{equation}
W_{\rm E} \leq {\eta}W_{\rm \frac{1}{2}} \leq 2{\eta}({\hat E} - E_{\rm low}) ,
\label{eqn:b3}
\end{equation}
where $\eta \approx$ 1.015, the maximum value of the ratio 
$W_{\rm E}/W_{\rm \frac{1}{2}}$.
A larger value of $W_{\rm \frac{1}{2}}$ 
would lead us to infer that there is a low-energy rollover in the spectrum, 
and not a line candidate.
Thus
\begin{equation}
p({\hat W_{\rm E}},{\hat E} \vert I) = \frac{\eta}{2({\hat E} - E_{\rm low})} .
\label{eqn:b4}
\end{equation}
The prior for a saturated line is simply the product
of eqs.~(\ref{eqn:b2}) and (\ref{eqn:b4}).
If the line is not saturated, we use eq.~(\ref{eqn:b3}) to
specify the prior for $W_{\rm \frac{1}{2}}$:
\begin{equation}
p({\hat W_{\rm \frac{1}{2}}} \vert {\hat W_{\rm E}},{\hat E},I) = \frac{\eta}{\hat W_{\rm \frac{1}{2}}} .
\label{eqn:b5}
\end{equation}
The product of eqs.~(\ref{eqn:b2}), (\ref{eqn:b4}), and (\ref{eqn:b5}) 
gives the prior for an unsaturated line.

Assignment of the priors for two lines follows a similar procedure.
We note three major differences:
\begin{itemize}
\item{if the model has harmonically spaced lines,
the prior range for $E_1$ is reduced:
$p({\hat E_1} \vert I) = (\frac{E_{\rm high}}{2} - E_{\rm low})^{-1}$;}
\item{the prior ranges for both
$W_{\rm E,1}$ and $W_{\rm E,2}$ are adapted to take into account the
fact that the two lines cannot overlap and be identified as two separate
lines;}
\item{and if the model has lines that are not harmonically spaced, then
$p({\hat E_2} \vert {\hat E_1})p({\hat E_1} \vert I) = [(E_{\rm high} - {\hat E_1})(E_{\rm high} - E_{\rm low})]^{-1}$.}
\end{itemize}

\newpage

\begin{figure}
\caption{
{\it Ginga} Proportional Counter (PC; top) and Scintillation
Counter (SC; bottom) time histories of GRB870303.  The PC data is
presented in 1 s bins, the SC data in 0.5 second bins.
The burst triggered the recording of {\it Ginga} burst-mode data
at $\approx$ 16 s; the preceding 16 s of burst-mode data,
in memory at the time of the trigger,
were recorded and not overwritten.
Burst-mode lasts for 64 seconds.  Epochs S1 (4 seconds)
and S2 (9 seconds) are shown;
the midpoints of S1 and S2 are separated by 22.5 seconds.}
\end{figure}

\begin{figure}
\caption{
{\it Ginga} GBD count-rate spectra for intervals S1 and S2 of GRB870303,
normalized by energy-loss bin width.}
\end{figure}

\begin{figure}
\caption{
(a): This flow chart illustrates how we select
the best-fit continuum model.  We begin by comparing the simplest
model ($M_1$) with all alternative models
that have a greater number of free parameters
($M_2$-$M_{\rm N}$), computing the significance of the decrease in
$s_{\rm m}^2$ for each ($\alpha_{\rm 1,2}$-$\alpha_{\rm 1,N}$).  
If no alternative model satisfies the criterion
$\alpha_{\rm \chi^2MLR} \leq$ 0.01, we select the simplest model;
otherwise, we select the simplest alternative model that fulfills the
criterion and repeat the comparison process, continuing until a continuum
model is selected (i.e., until no alternative model satisfies the 
criterion).
(b): This flow chart illustrates how we
select the range of usable PC bins.  
We do not use all PC data because of the difficulty of modeling
the spectral rollover at energy-losses $\simless$ 5 keV.  
The box ``Select Continuum Model'' refers to the flow chart in (a).
}
\end{figure}

\begin{figure}
\caption{
Probability contours resulting from the use of
various combinations of unnormalized
Gaussian amplitude and width ($\beta$,$\sigma$), equivalent width 
$W_{\rm E}$, and full-width $W_{\rm \frac{1}{2}}$, to
parametrize line shape.
We show 1, 2, and 3$\sigma$
contours, representing 68.3\%, 95.5\%, and 99.7\% of the integrated
probability.
Only the $(W_{\rm \frac{1}{2}},W_{\rm E})$ parametrization contours, 
shown at lower left, show the
elliptical behavior required to use eq.~(\ref{eqn:odds}) to compute the
Bayesian odds.}
\end{figure}

\begin{figure}
\caption{
The ratio of line equivalent width, $W_{\rm E}$, to line
full-width, $W_{\rm \frac{1}{2}}$, as a function of Gaussian amplitude
$\beta$.  This ratio is not a function of the Gaussian width
$\sigma$.  It reaches 1 (short dashed line) when
$\beta \approx$ 4.75 and peaks at $r \approx$ 1.015
when $\beta$ = $\beta_o \approx$ 18.75 (long dashed line).
We set $\beta = \beta_o$ when fitting saturated lines to data.}
\end{figure}

\begin{figure}
\caption{
Values of the fitting statistics $s_{\rm m}^2$ (squares) and $L$ (circles) as
a function of $\cos{\theta_{\rm inc}}$, for fits to the data of
GRB870303 S1 (top) and GRB870303 S2 (bottom).  Unfilled shapes
represent continuum model fits, while filled shapes represent 
continuum-plus-line(s) model fits.  Any jaggedness in
fitting statistic values as a function of angle reflects the use of
Monte Carlo simulations to create {\it Ginga} GBD response matrices.
Jaggedness is more apparent in fits to the data of S2 because they
are more informative (i.e.~the number of counts per bin is higher for
S2 than S1).
}
\end{figure}

\begin{figure}
\caption{
Best-fit continuum-plus-line(s) photon number spectra (top),
observed count-rate spectra and best-fit continuum-plus-line(s) count-rate
spectra (middle), 
and residuals of the best-fit in units of $\sigma$ (bottom) for the
intervals S1 (left) and S2 (right) 
of GRB870303.
}
\end{figure}

\begin{figure}
\caption{
Same as Figure 6, for the joint fits to intervals
S1 and S2 of GRB870303.
}
\end{figure}

\begin{figure}
\caption{
Best-fit continuum-plus-line(s) photon number spectra (top),
observed count-rate spectra and best-fit continuum-plus-line(s) count-rate
spectra (middle), 
and residuals of the best-fit in units of $\sigma$ (bottom) for the
joint fit to intervals
S1 (left) and S2 (right) of GRB870303.
}
\end{figure}

\begin{figure}
\caption{
Best-fit photon spectra for GRB870303 S1,
assuming burst photon incidence angle $\theta_{\rm inc}$ = 37.7$^{\circ}$.
The solid line shows the best-fit PL continuum model,
while the dashed line shows the best-fit Band et al.~continuum model.
}
\end{figure}

\clearpage

\begin{deluxetable}{lc}
\tablenum{1}
\tablecaption{Continuum Models}
\tablewidth{0pt}
\tablehead{
\colhead{Model} & \colhead{Formula} }
\startdata
Power Law (PL) & $\frac{dN}{dE}=A E^{-\alpha}$ \\
PL $\times$ Exponential (PLE) & $\frac{dN}{dE}=A E^{-\alpha} \exp(-\frac{E}{E_c})$ \\
\\
Band et al.~(1993) & $\frac{dN}{dE}=\cases{ A E^{-\alpha_1} \exp(-\frac{E}{E_{\rm c}}),&$E \leq (\alpha_2-\alpha_1)E_{\rm c}$\cr A \left[(\alpha_2-\alpha_1)E_{\rm c}\right]^{\alpha_2-\alpha_1}~\times \cr~~~~~~~~\exp(\alpha_1-\alpha_2) E^{-\alpha_2} ,&otherwise\cr}$ \\
\\
Broken Power Law (BPL) & $\frac{dN}{dE}=\cases{ A E^{-\alpha_1},&$E \leq E_b$\cr A E_b^{\alpha_2-\alpha_1} E^{-\alpha_2},&otherwise\cr}$ \\
\\
Low-Energy Rollover (NH) & $\left(\frac{dN}{dE}\right)_{\rm col}=\left(\frac{dN}{dE}\right) \exp(-A \times E^{-3})$ \\
\enddata
\tablecomments{In the fitting code itself, it is the logarithms of the normalization $A$ and PLE cutoff energy $E_c$ that are varied.  Also, a ``pivot" energy of 20 keV is used to reduce the size of the frequentist confidence and Bayesian credible intervals for the highly correlated parameters $A$ and $\alpha$.  These changes alter the likelihood surface in parameter space in such a way as to make it more closely resemble a multi-dimensional Gaussian function.}
\end{deluxetable}

\clearpage

\begin{deluxetable}{cccc}
\tablenum{2}
\tablecaption{Energy-Loss Bins and Continua Used in Analyses}
\tablewidth{0pt}
\tablehead{
\colhead{Spectrum} & \colhead{PC Bins} & \colhead{SC Bins} & \colhead{Selected Continuum} }
\startdata
S1    & 8-15 & 2-31 & PL \\
\\
S2    & 7-15 & 2-31 & PLE \\
\\
S1+S2 & 10-15 & 2-31 & PL(S1)+PLE(S2)\tablenotemark{a} \\
\enddata
\tablenotetext{a}{The continuum model is a six-parameter model in which ${\log}A_{\rm S1}$ and $\alpha_{\rm S1}$ vary independently of ${\log}A_{\rm S2}$ and $\alpha_{\rm S2}$, but for which $\cos(\theta_{\rm inc})_{\rm S1} = \cos(\theta_{\rm inc})_{\rm S2}$.}
\end{deluxetable}

\clearpage

\begin{deluxetable}{cc}
\tablenum{3}
\tablecaption{Exponentiated Gaussian Line Models}
\tablewidth{0pt}
\tablehead{\colhead{~~~~~~Model~~~~~~} & \colhead{~~~Free Line Parameters~~~} }
\startdata
S1-S & $(E,W_{\rm E})$ \\
S1-U & $(E,W_{\rm E},W_{\rm \frac{1}{2}})$ \\
\\
S2-A & $(E_1,W_{\rm E,1})$ \\
S2-B & $(E_1,W_{\rm E,1},W_{\rm E,2})$ \\
S2-C & $(E_1,W_{\rm E,1},W_{\rm \frac{1}{2},1})$ \\
S2-D & $(E_1,W_{\rm E,1},W_{\rm \frac{1}{2},1},W_{\rm E,2},W_{\rm \frac{1}{2},2})$ \\
\enddata
\tablecomments{Parameters not shown have values set by the values of
the parameters shown; e.g., for S2-A, $E_2 = 2E_1$, $W_{\rm \frac{1}{2},1} =
{\eta}^{-1} W_{\rm E,1}$, $W_{\rm E,2} = 2W_{\rm E,1}$, and 
$W_{\rm \frac{1}{2},2} = {\eta}^{-1} W_{\rm E,2}$.
Not shown are variations on the S2 class of models for
which $E_2 \neq 2E_1$.  While these models were tested for completeness,
none significantly improved fits.}
\end{deluxetable}

\clearpage

\begin{deluxetable}{cl}
\tablenum{4}
\tablecaption{Prior Probabilities for Exponentiated Gaussian Model}
\tablewidth{0pt}
\tablehead{
\colhead{Model} & \colhead{Prior Probability} }
\startdata
S1-S & $p~=~[2{\eta}({\hat E}-E_{\rm low})(E_{\rm high}-E_{\rm low})]^{-1}$\\
S1-U & $p~=~[2{\hat W}_{\rm E}({\hat E}-E_{\rm low})(E_{\rm high}-E_{\rm low})]^{-1}$\\
\\
S2-A & $p~=~[\frac{\eta}{3}{\hat E}_1(E_{\rm high}-2E_{\rm low})]^{-1}$\\
S2-B & $p~=~[(2{\eta}{\hat E}_1-{\hat W}_{\rm E,1})({\hat E}_1-E_{\rm low})(E_{\rm high}-2E_{\rm low})]^{-1}$\\
S2-C & $p~=~[\frac{\eta}{3}{\hat W}_{\rm \frac{1}{2},1}{\hat E}_1(E_{\rm high}-2E_{\rm low})]^{-1}$\\
S2-D & $p~=~[{\eta}^2{\hat W}_{\rm \frac{1}{2},1}{\hat W}_{\rm \frac{1}{2},2}(2{\hat E}_1-{\hat W}_{\rm \frac{1}{2},1})({\hat E}_1-E_{\rm low})(E_{\rm high}-2E_{\rm low})]^{-1}$\\
\enddata
\tablecomments{$\eta~=~\left( 
{{W_E}\over{W_{\frac{1}{2}}}} \right)_{max}~\approx$ 1.015. 
$E_{\rm low}$ and $E_{\rm high}$ represent the low and high energy-loss 
bandpass boundaries, respectively, for the {\it Ginga} GBD.  
Not shown are the priors for 
variations on the S2 class of models for
which $E_2 \neq 2E_1$; while these models were tested,
none significantly improved fits.  See Appendix B for details.}
\end{deluxetable}

\clearpage

\begin{deluxetable}{cccccccc}
\tablenum{5}
\tablecaption{Line Significances and Odds}
\tablewidth{0pt}
\tablehead{
\colhead{Dataset} & \colhead{Model} & \colhead{$s_{\rm C}^2$~(dof)} & \colhead{$s_{\rm C+L}^2$~(dof)} & \colhead{$\alpha_{\rm \chi^2 MLR}$} & \colhead{$L_{\rm C}$} & \colhead{$L_{\rm C+L}$} & \colhead{Odds} }
\startdata
S1 & S1-S & 42.71~(35) & 22.22~(33) & 3.6 $\times$ 10$^{-5}$ & 49.44 & 60.55 & 114:1 \\
S2 & S2-A & 49.94~(35) & 32.60~(33) & 1.7 $\times$ 10$^{-4}$ & 89.90 & 98.73 & 7:1 \\
S1+S2 & S2-B\tablenotemark{a} & 86.18~(66) & 46.13~(62) & 4.2 $\times$ 10$^{-8}$ & 126.44 & 147.28 & 40,300:1 \\
\enddata
\tablenotetext{a}{All line parameters have the same value for S1 and S2 except $W_{\rm E,1,S1} \neq W_{\rm E,1,S2}$.}
\end{deluxetable}

\clearpage

\begin{deluxetable}{ccccccc}
\tablenum{6}
\tablecaption{Best-Fit Parameters}
\tablewidth{0pt}
\tablehead{
 & & \colhead{Frequentist} & & & \colhead{Bayesian} & \\
\colhead{Parameter} & \colhead{S1} & \colhead{S2} & \colhead{S1+S2} & \colhead{S1} & \colhead{S2} & \colhead{S1+S2} }
\startdata
${\log}A_{\rm S1}$\tablenotemark{a} & -0.77 & - & -0.80 & -0.84 & - & -0.80\\
${\log}A_{\rm S2}$\tablenotemark{a} & - & -0.56 & -0.39 & - & -0.55 & -0.39\\
$\alpha_{\rm S1}$ & 1.72 & - & 1.75 & 1.67 & - & 1.76\\
$\alpha_{\rm S2}$ & - & 1.19 & 1.48 & - & 1.19 & 1.47\\
$E_c$ (keV) & - & 2.14 & 2.39 & - & 2.13 & 2.38\\
\\
$E_1$ (keV) & 21.4 & 21.8 & 21.5 & 21.3 & 21.8 & 21.5\\
$W_{E,1,{\rm S1}}$ (keV) & 10.7 & - & 11.2 & 10.4 & - & 11.1\\
$W_{E,1,{\rm S2}}$ (keV) & - & 2.16 & 2.82 & - & 2.21 & 2.73\\
$W_{E,2}$ (keV) & - & - & 2.72 & - & - & 2.86\\
\\
${\cos}(\theta_{\rm inc})$ & 0.54 & 0.82 & 0.60 & 0.58 & 0.82 & 0.60\\
\enddata
\tablenotetext{a}{Amplitudes at the ``pivot" energy of 20 keV.}
\end{deluxetable}

\clearpage

\begin{deluxetable}{lcccccc}
\tablenum{7}
\tablecaption{S1: Parameter Estimation}
\tablewidth{0pt}
\tablehead{
 & \colhead{Par} & \colhead{Freq} & \colhead{Bayes} &
\colhead{Par} & \colhead{Freq} & \colhead{Bayes}
}
\startdata
Best Fit~~~~~& ${\log}A$\tablenotemark{a} & -0.77 & -0.84 & $E$ & 21.4 & 21.3\\
1$\sigma$  &  & [-1.05,-0.74] & [-1.14,-0.84]& (keV) & [20.3,22.5] & [20.0,22.5]\\
2$\sigma$  &  & [-1.23,-0.70] & [-1.23,-0.73]&  & [19.2,23.9] & [18.7,24.0]\\
3$\sigma$  &  & [-1.29,-0.67] & [-1.31,-0.66]&  & [18.1,25.7] & [17.5,25.8]\\
Best Fit   & $\alpha$  & 1.72 & 1.67 & $W_E$ & 10.7 & 10.4 \\
1$\sigma$  &  & [1.52,1.78] & [1.47,1.70]& (keV) & [8.41,13.1] & [7.85,12.9]\\
2$\sigma$  &  & [1.35,1.85] & [1.37,1.80]&  & [6.16,15.9] & [5.74,15.8] \\
3$\sigma$  &  & [1.26,1.92] & [1.27,1.90]&  & [3.80,19.8] & [4.55,17.8] \\
Best Fit   & & & & ~~~$\cos(\theta_{\rm inc})$~~~ & 0.54 & 0.58 \\
1$\sigma$  & & & &          & [0.54,0.78] & [0.54,0.72]\\
2$\sigma$  & & & &          & [0.54,0.98] & [0.54,0.95]\\
3$\sigma$  & & & & 	  & [0.54,0.98] & [0.54,0.978]\\
\enddata
\tablenotetext{a}{Amplitude at the ``pivot" energy of 20 keV.}
\end{deluxetable}

\clearpage

\begin{deluxetable}{lcccccc}
\tablenum{8}
\tablecaption{S2: Parameter Estimation}
\tablewidth{0pt}
\tablehead{
 & \colhead{Par} & \colhead{Freq} & \colhead{Bayes} &
\colhead{Par} & \colhead{Freq} & \colhead{Bayes}
}
\startdata
Best Fit~~~~& ${\log}A$\tablenotemark{a} & -0.56 & -0.55 & $E_1$ & 21.8 & 21.8\\
1$\sigma$  &  & [-0.60,-0.43] & [-0.63,-0.47]& (keV) & [20.3,22.5] & [21.1,22.6]\\
2$\sigma$  &  & [-0.69,-0.34] & [-0.70,-0.38]&  & [19.2,23.9] & [20.2,23.3]\\
3$\sigma$  &  & [-0.72,-0.28] & [-0.73,-0.34]&  & [18.1,25.7] & [19.5,24.0]\\
Best Fit   & $\alpha$  & 1.19 & 1.19 & $W_{E,1}$ & 2.16 & 2.21 \\
1$\sigma$  &  & [1.10,1.32] & [1.15,1.31]& (keV) & [1.73,2.62] & [1.67,2.70]\\
2$\sigma$  &  & [0.97,1.44] & [1.00,1.42]&  & [1.20,3.15] & [1.16,3.09] \\
3$\sigma$  &  & [0.89,1.54] & [0.90,1.54]&  & [0.63,3.66] & [1.02,3.38] \\
Best Fit   & ${\log}E_c$ & 2.14 & 2.13 & ~~$\cos(\theta_{\rm inc})$~~ & 0.82 & 0.82 \\
1$\sigma$  & (keV) & [2.04,2.25] & [2.04,2.25] &     & [0.70,0.94] & [0.66,0.96]\\
2$\sigma$  &  & [1.96,2.41] & [1.94,2.39] &     & [0.58,0.98] & [0.60,0.98]\\
3$\sigma$  &  & [1.88,2.58] & [1.84,2.57] &     & [0.54,0.98] & [0.56,0.98]\\
\enddata
\tablenotetext{a}{Amplitude at the ``pivot" energy of 20 keV.}
\end{deluxetable}

\clearpage  

\begin{deluxetable}{lcccc}
\tablenum{9}
\tablecaption{S1+S2: Parameter Estimation}
\tablewidth{0pt}
\tablehead{
 & \colhead{Par} & \colhead{Freq} &
\colhead{Par} & \colhead{Freq} }
\startdata
Best Fit~~~~& ${\log}A_{\rm S1}$\tablenotemark{a} & -0.80 & $E_1$ & 21.5 \\
1$\sigma$  &  & [-0.98,-0.73] & (keV) & [20.8,22.1] \\
2$\sigma$  &  & [-1.11,-0.67] & & [20.1,22.6] \\
3$\sigma$  &  & [-1.22,-0.61] & & [19.3,23.2] \\
Best Fit   & $\alpha_{\rm S1}$  & 1.75 & $W_{E,1,{\rm S1}}$ & 11.2 \\
1$\sigma$  &  & [1.61,1.83] & (keV) & [9.15,13.4] \\
2$\sigma$  &  & [1.49,1.92] &  & [7.08,16.0] \\
3$\sigma$  &  & [1.37,2.03] &  & [5.02,19.2] \\
Best Fit   & ${\log}A_{\rm S2}$\tablenotemark{a} & -0.39 & $W_{E,1,{\rm S2}}$ & 2.82 \\
1$\sigma$  &  & [-0.43,-0.33] & (keV) & [2.04,3.40] \\
2$\sigma$  &  & [-0.44,-0.24] & & [1.35,4.00] \\
3$\sigma$  &  & [-0.47,-0.22] & & [0.64,4.71] \\
Best Fit   & $\alpha_{\rm S2}$  & 1.48 & $W_{E,2}$ & 2.72 \\
1$\sigma$  &  & [1.30,1.56] & (keV) & [1.38,4.06] \\
2$\sigma$  &  & [1.12,1.67] &  & [0.00,5.28] \\
3$\sigma$  &  & [0.98,1.74] &  & [0.00,6.44] \\
Best Fit   & ${\log}E_c$ & 2.40 & ~~$\cos(\theta_{\rm inc})$~~ & 0.60 \\
1$\sigma$  & (keV) & [2.22,2.59] & & [0.58,0.78] \\
2$\sigma$  &  & [2.08,2.91] &     & [0.54,0.96] \\
3$\sigma$  &  & [1.97,$\infty$] & & [0.54,0.98] \\
\enddata
\tablenotetext{a}{Amplitude at the ``pivot" energy of 20 keV.}
\end{deluxetable}

\clearpage  

\begin{deluxetable}{ccccccccc}
\tablenum{10}
\tablecaption{Analyses of GRB870303 Line Candidates}
\tablewidth{0pt}
\tablehead{
 & & & & \colhead{Model} & \colhead{C} & \colhead{C+L} & \colhead{Number} & \colhead{Sig} \\ 
 & & & & \colhead{Comp} & \colhead{Model} & \colhead{Model} & \colhead{of Line} & \colhead{or} \\ 
\colhead{Work} & \colhead{Spec} & $\theta_{\rm inc}$ & \colhead{Stat} & \colhead{Test} & \colhead{Comp?} & \colhead{Comp?} & \colhead{Par} & \colhead{Odds}
}
\startdata
M88 & S2 & 37.7$^{\circ}$ & $s_{\rm d}^2$ & $F$ & N\tablenotemark{a} & N & 6 & $\sim$ 10$^{-3}$ \\
G92 & S1 & 37.7$^{\circ}$ & $s_{\rm d}^2$ & ${\chi^2}$~MLR & Y & N & 3 & 1$\times$10$^{-5}$ \\
    & S2 &     &         &                & Y & N & 6 & 2$\times$10$^{-4}$ \\
G93 & S1 & 37.7$^{\circ}$ & $L$ & Odds & N\tablenotemark{b} & N & 2\tablenotemark{c} & 110:1 \\
    & S2 &  &   &      & N\tablenotemark{d} & N & 4\tablenotemark{e}  & 2.8:1 \\
\hline
This& S1 & Free & $s_{\rm m}^2$ & ${\chi^2}$~MLR & Y & Y & 2 & 2.2$\times$10$^{-5}$ \\
Work&      &                         & $L$ &   Odds & N  & Y   & 2 & 114:1 \\
                       & S2 & & $s_{\rm m}^2$ & ${\chi^2}$~MLR & Y   & Y   & 2 & 1.7$\times$10$^{-4}$ \\
                       &      &                         & $L$ &   Odds & N   & Y   & 2 & 7:1 \\
                       & S1+S2 & & $s_{\rm m}^2$ & ${\chi^2}$~MLR & Y   & Y   & 4 & 4.2$\times$10$^{-8}$ \\
                       &        &                           & $L$ &   Odds & N   & Y   & 4 & 40,300:1 \\
\enddata
\tablenotetext{a}{M88 assume a thermal cyclotron continuum.}
\tablenotetext{b}{G93 apply the power-law continuum model used in G92.}
\tablenotetext{c}{G93 assume the line to be saturated.}
\tablenotetext{d}{G93 apply the two-segment broken power-law continuum model used in G92.}
\tablenotetext{e}{G93 assume the energies and full-widths of the two 
lines to be harmonically related.}
\end{deluxetable}

\clearpage  

\begin{deluxetable}{cccccc}
\tablenum{11}
\tablecaption{Effect of Changing Continuum Model and PC Bin Range on Fits}
\tablewidth{0pt}
\tablehead{
\colhead{Dataset} & \colhead{Continuum} & \colhead{PC Bins} & \colhead{$\alpha_{\rm \chi^2MLR}$} & Odds }
\startdata
S1 & PL & 8-15 & 1.2$\times$10$^{-5}$ & 119:1 \\
   & PL & 10-14 & 5.3$\times$10$^{-6}$ & 339:1 \\
   & Band & 8-15 & 1.6$\times$10$^{-6}$ &  -\tablenotemark{a}\\
   & Band & 10-14 & 3.9$\times$10$^{-6}$ & -\tablenotemark{a} \\
S2 & PLE & 7-15 & 4.0$\times$10$^{-5}$ & 14:1 \\
   & PLE & 10-14 & 1.1$\times$10$^{-5}$ & 15:1 \\
   & Band & 7-15 & 1.1$\times$10$^{-5}$ & 81:1 \\
   & Band & 10-14 & 4.0$\times$10$^{-6}$ & 87:1 \\
   & BPL & 7-15 & 2.0$\times$10$^{-5}$ & 28:1 \\
   & BPL & 10-14 & 5.3$\times$10$^{-5}$ & 10:1 \\
\enddata
\tablecomments{We use the best-fit line model for each dataset, 
and assume $\theta_{\rm inc}$ = 37.7$^{\circ}$.}
\tablenotetext{a}{$\sqrt{{\det}V}$ cannot be computed because the mode is
too close to a parameter space boundary.}
\end{deluxetable}

\clearpage  

\begin{deluxetable}{cccccc}
\tablenum{12}
\tablecaption{Effect of Increasing the Number of Line Parameters in Fits to S1 and S2}
\tablewidth{0pt}
\tablehead{
\colhead{} & \colhead{2 Par (S1-S)} & \colhead{3 Par (S1-U)} & \colhead{2 Par (S2-A)} & \colhead{5 Par (S2-D)} & \colhead{6 Par} }
\startdata
$s_{\rm m,C}^2$ & 44.84 & 44.84 & 53.91 & 53.91 & 53.91 \\
$s_{\rm m,C+L}^2$ & 22.23 & 22.23 & 33.63 & 31.13 & 30.49 \\
$\alpha_{\rm \chi^2MLR}$ & 1.2$\times$10$^{-5}$ & 4.8$\times$10$^{-5}$ & 4.0$\times$10$^{-5}$ & 3.7$\times$10$^{-4}$ & 6.7$\times$10$^{-4}$ \\
\\
$E_1$ (keV) & 21.2 & 21.2 & 21.8 & 21.8 & 21.3 \\
$W_{\rm E,1}$ (keV) & 9.98 & 9.98 & 2.24 & 3.35 & 3.12 \\
$W_{\rm \frac{1}{2},1}$ (keV) & [10.1] & 10.1 & [2.27] & 6.20 & 5.55 \\
$E_2$ (keV) & & & [43.6] & [43.6] & 44.4 \\
$W_{\rm E,2}$ (keV) & & & [5.48] & 4.23 & 4.25 \\
$W_{\rm \frac{1}{2},2}$ (keV) & & & [5.54] & 4.67 & 4.80 \\
\enddata
\tablecomments{Values in brackets are fixed by values of other parameters
(see Table 3).  We assume $\theta_{\rm inc}$ = 37.7$^{\circ}$.}
\end{deluxetable}

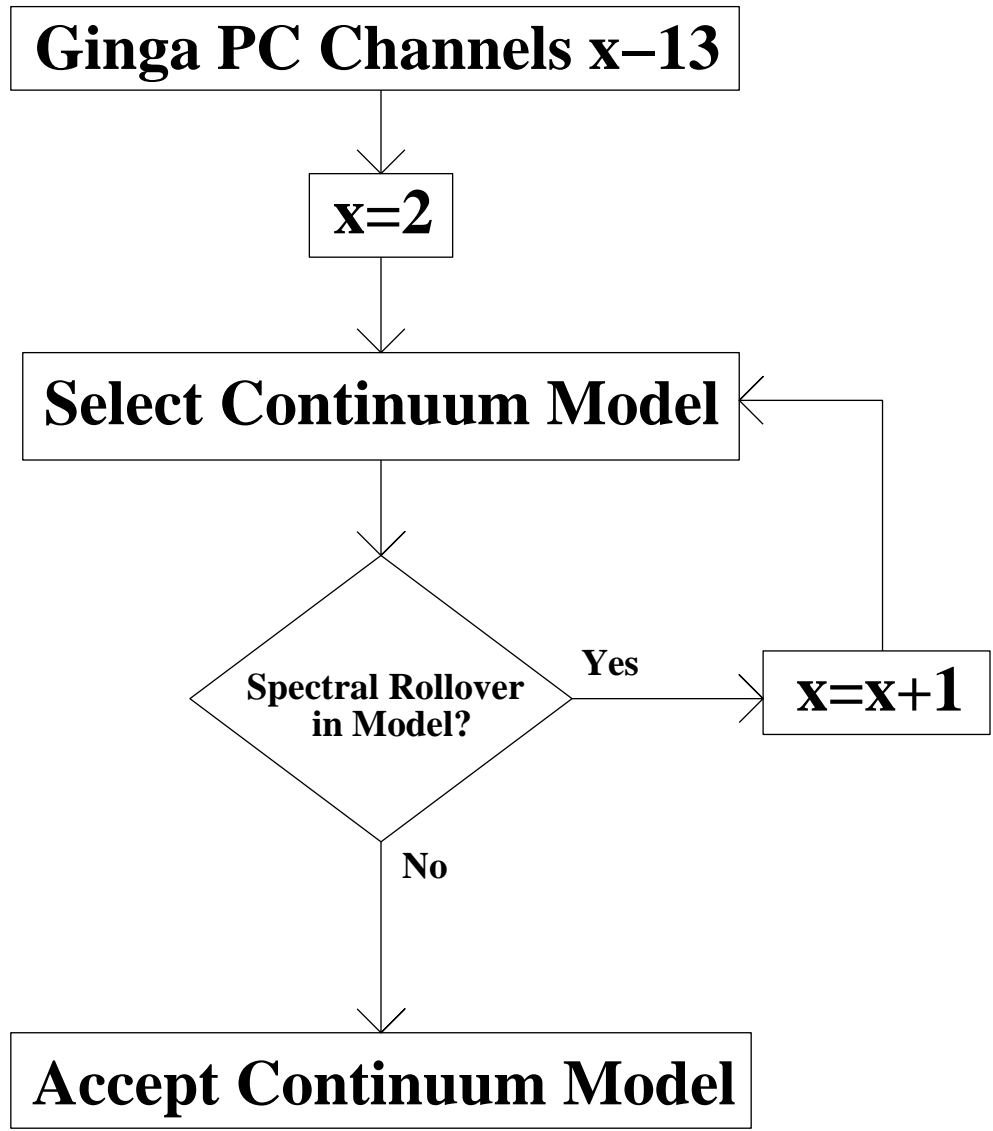
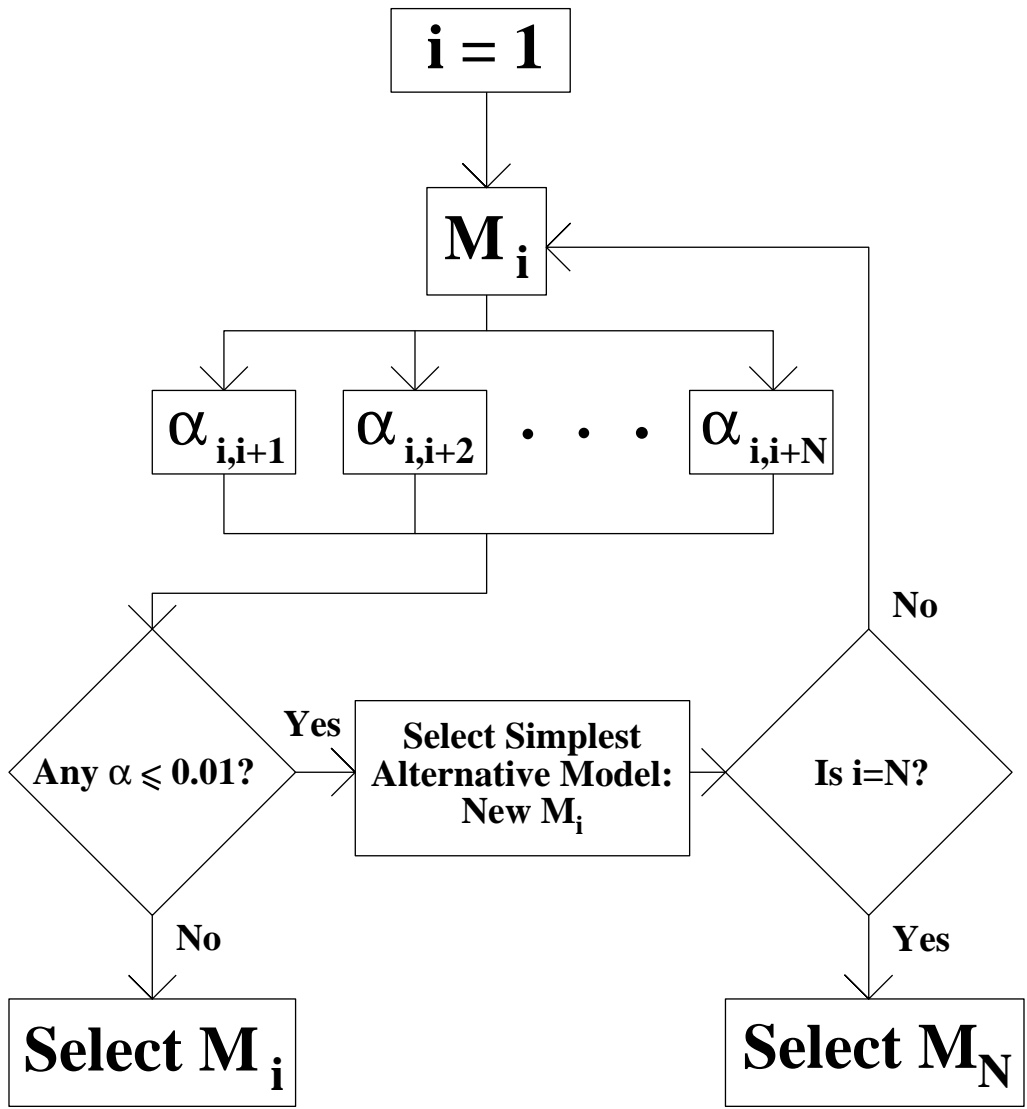
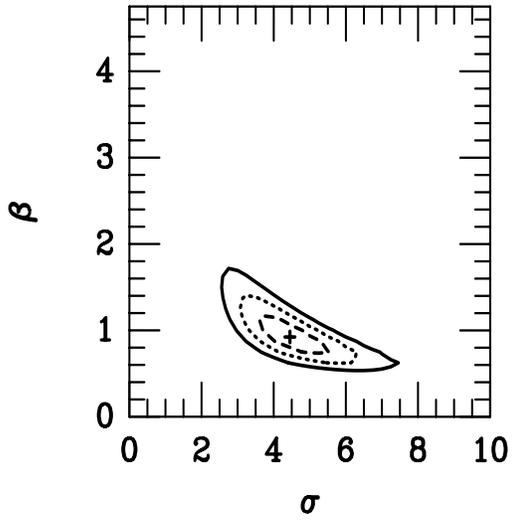
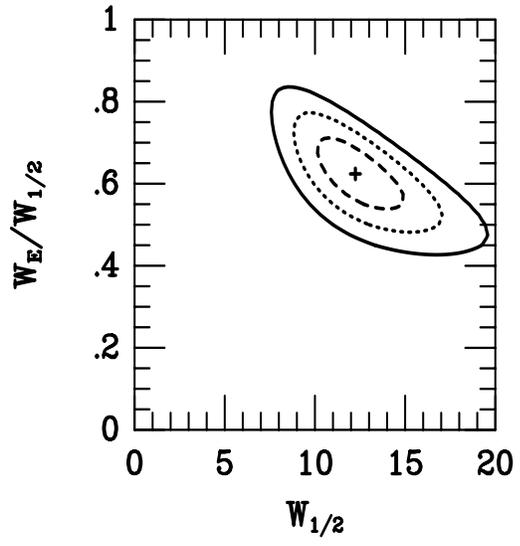
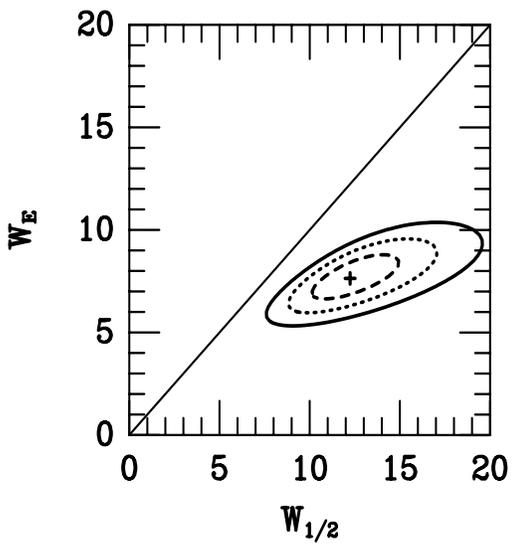
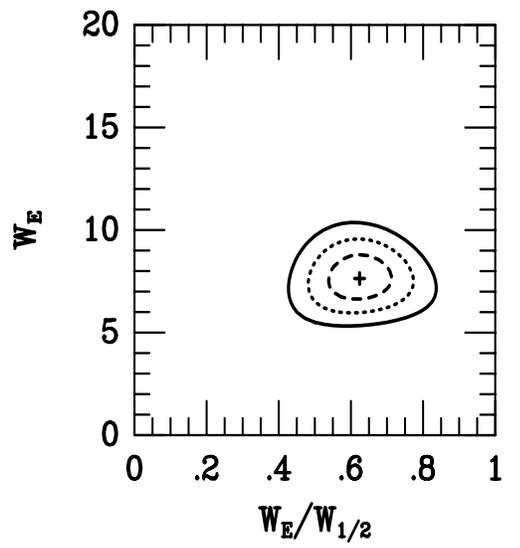
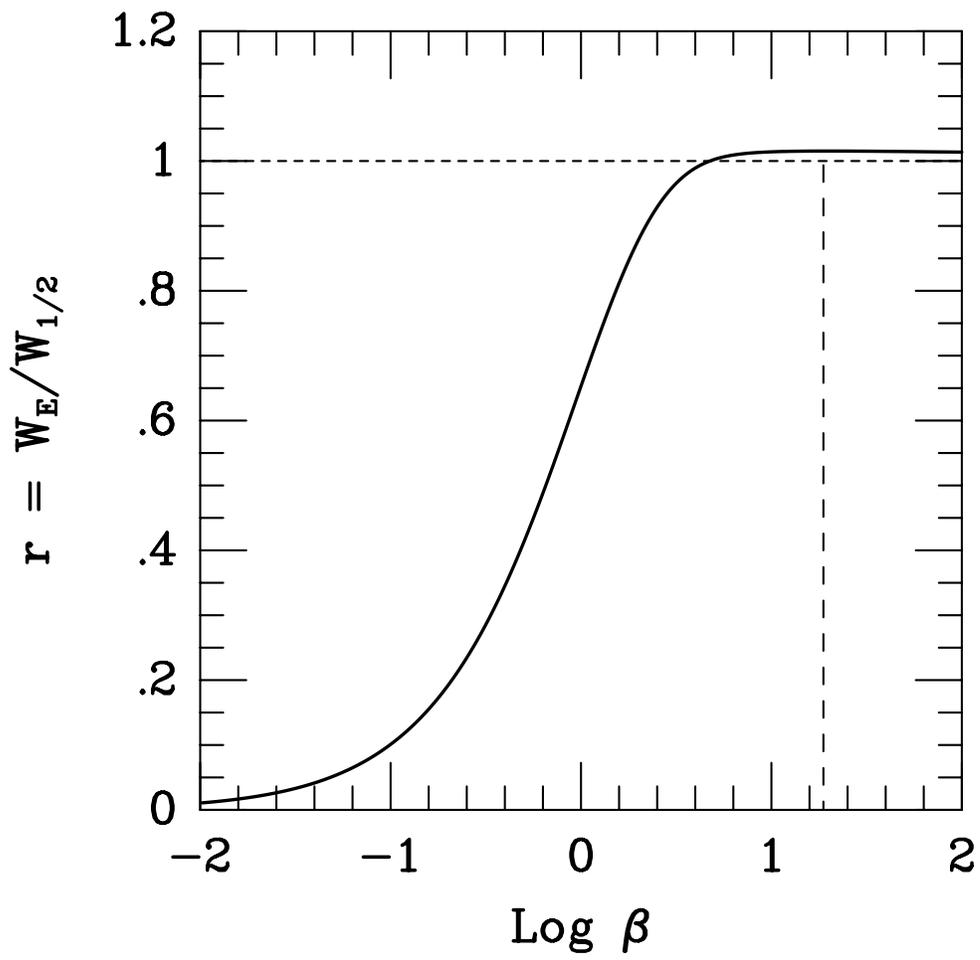
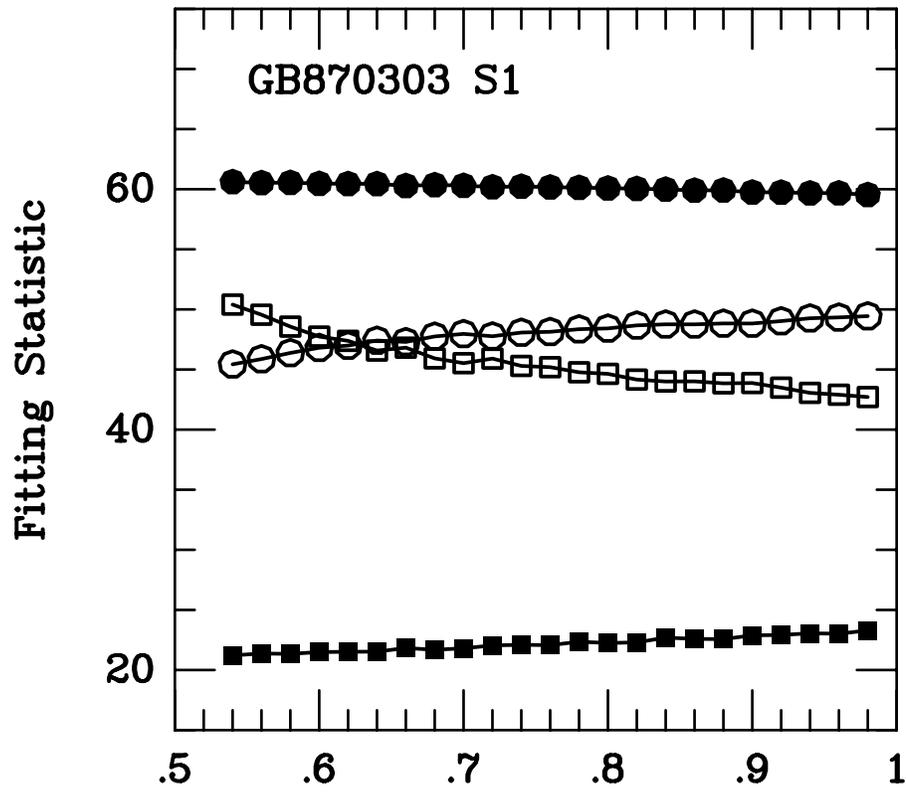
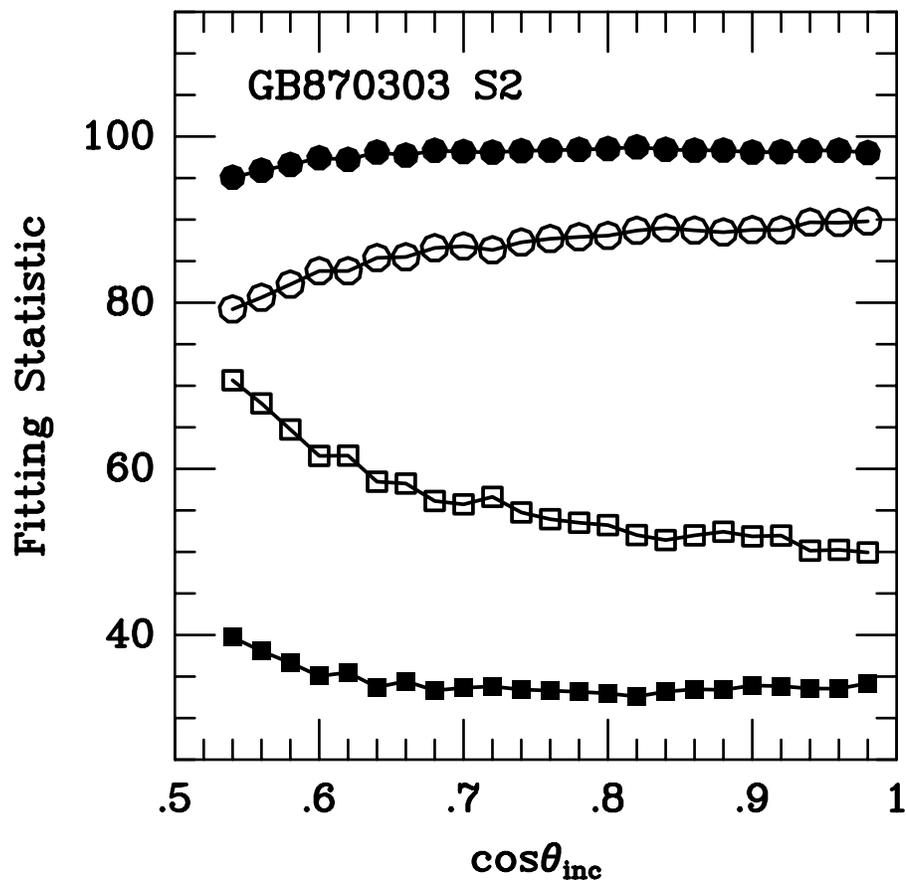
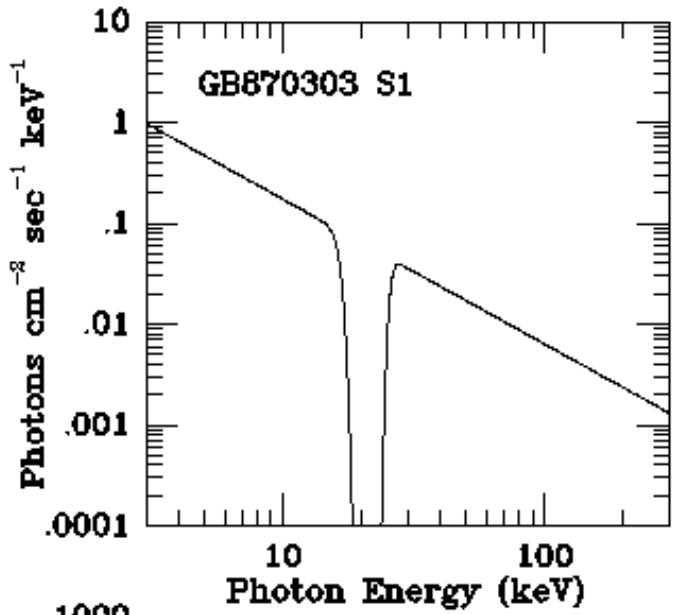
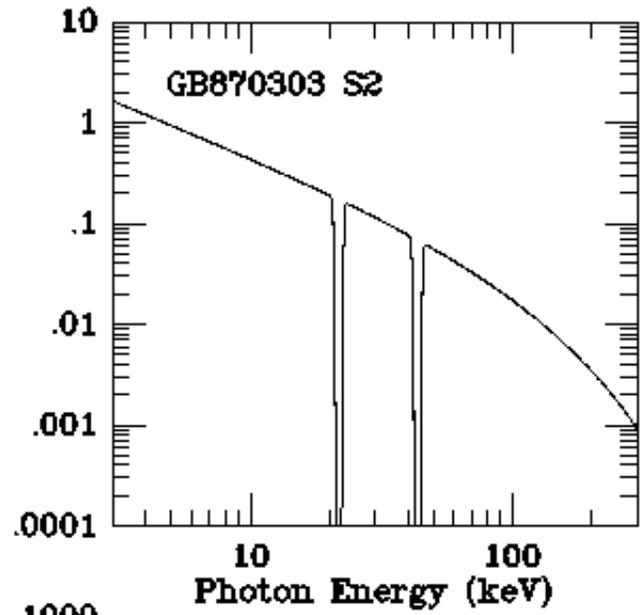
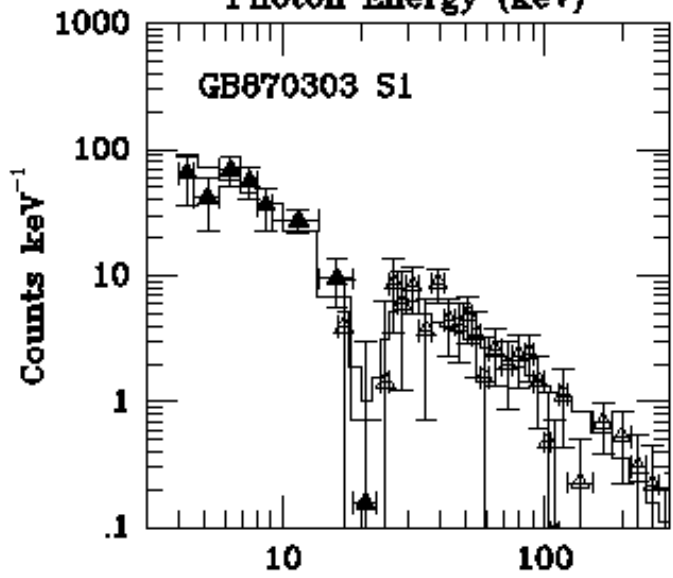
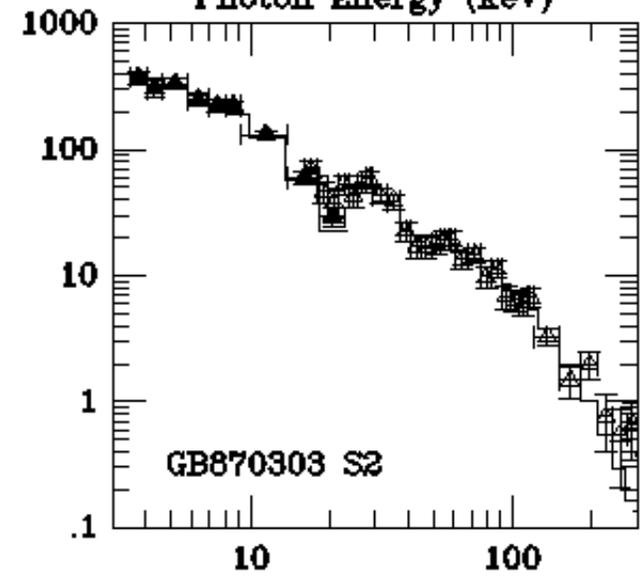
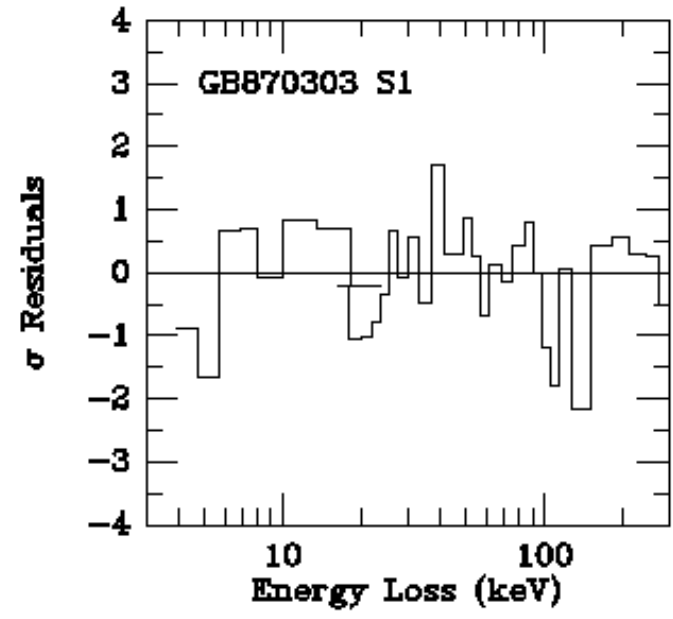
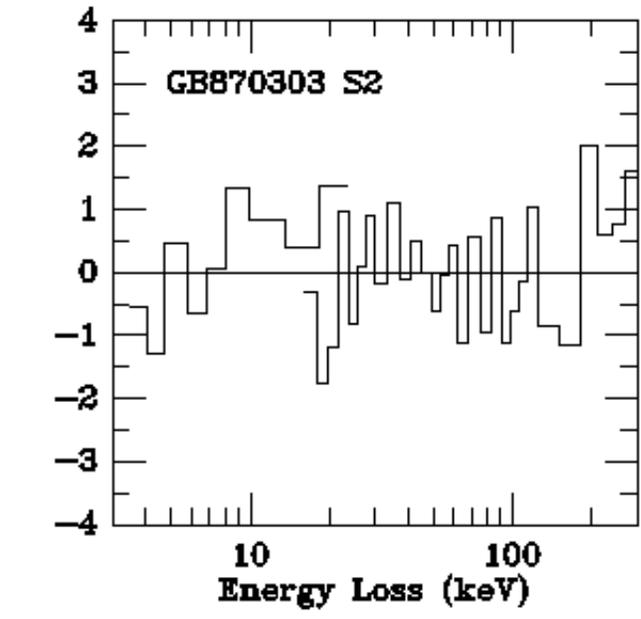
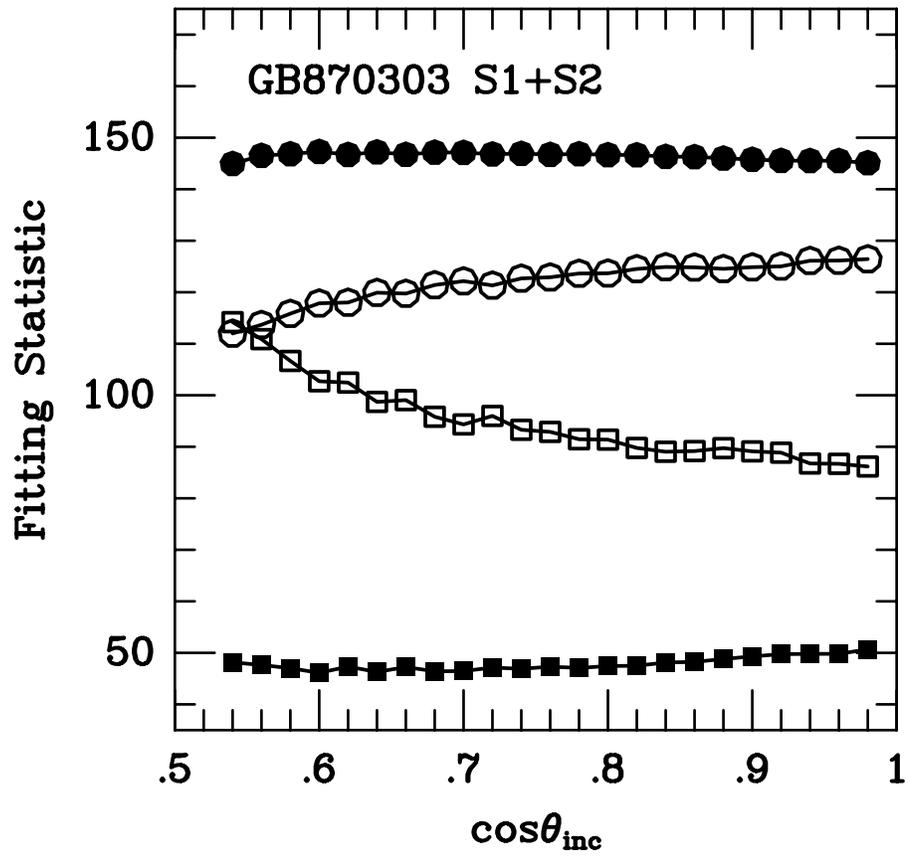
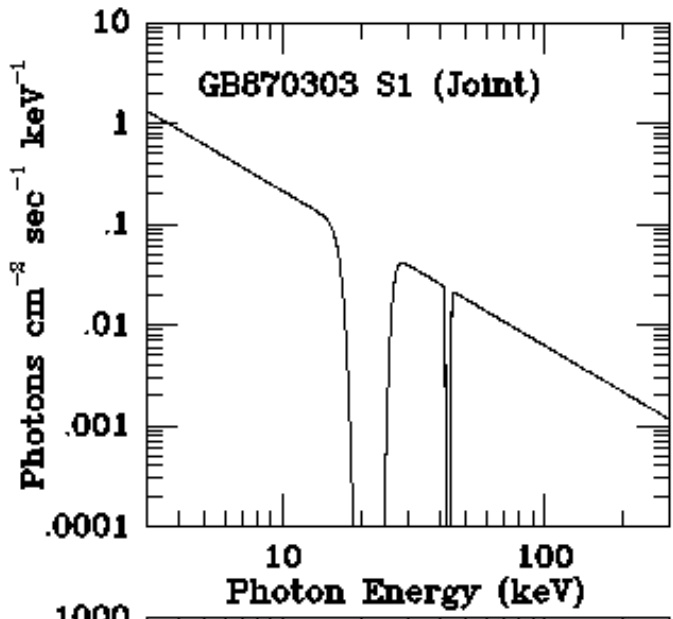
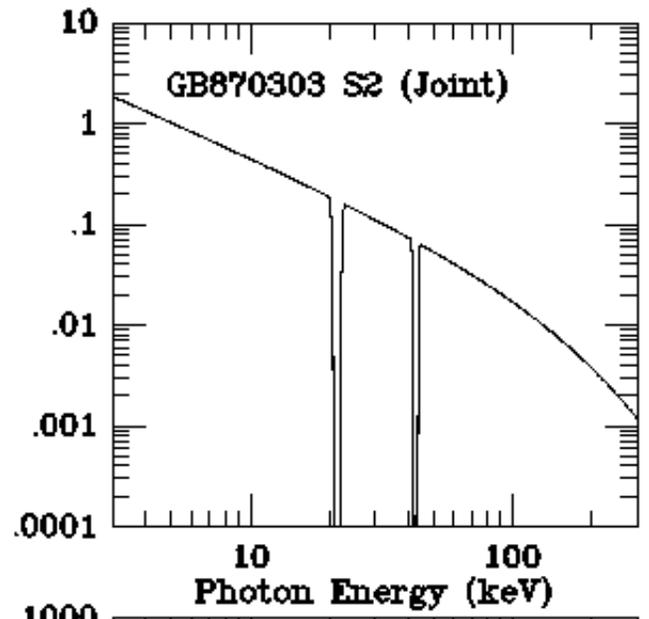
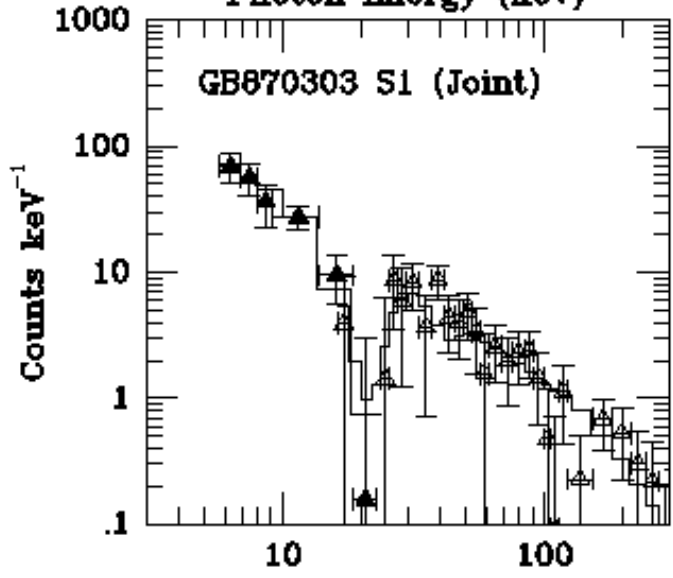
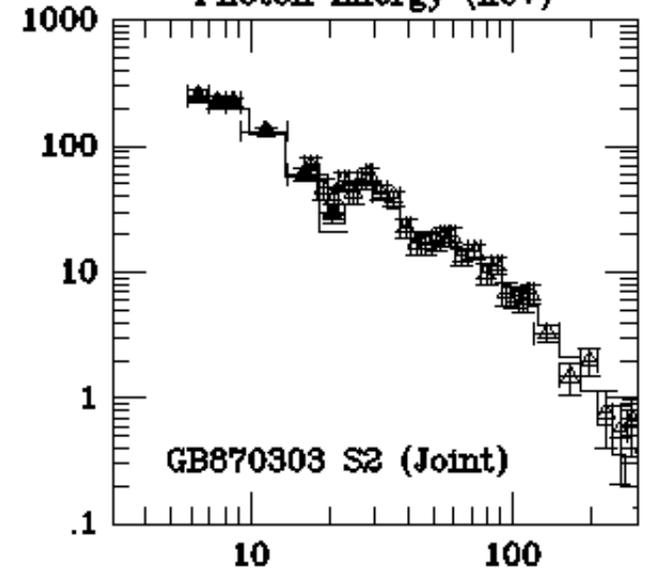
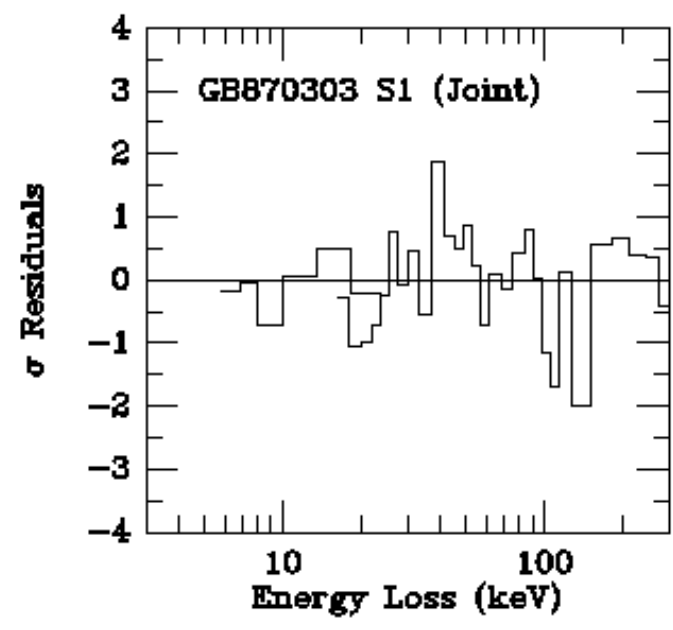
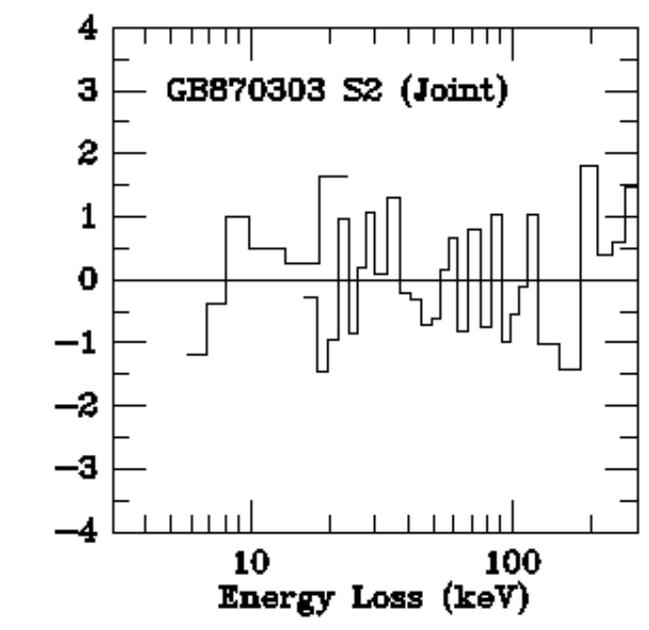
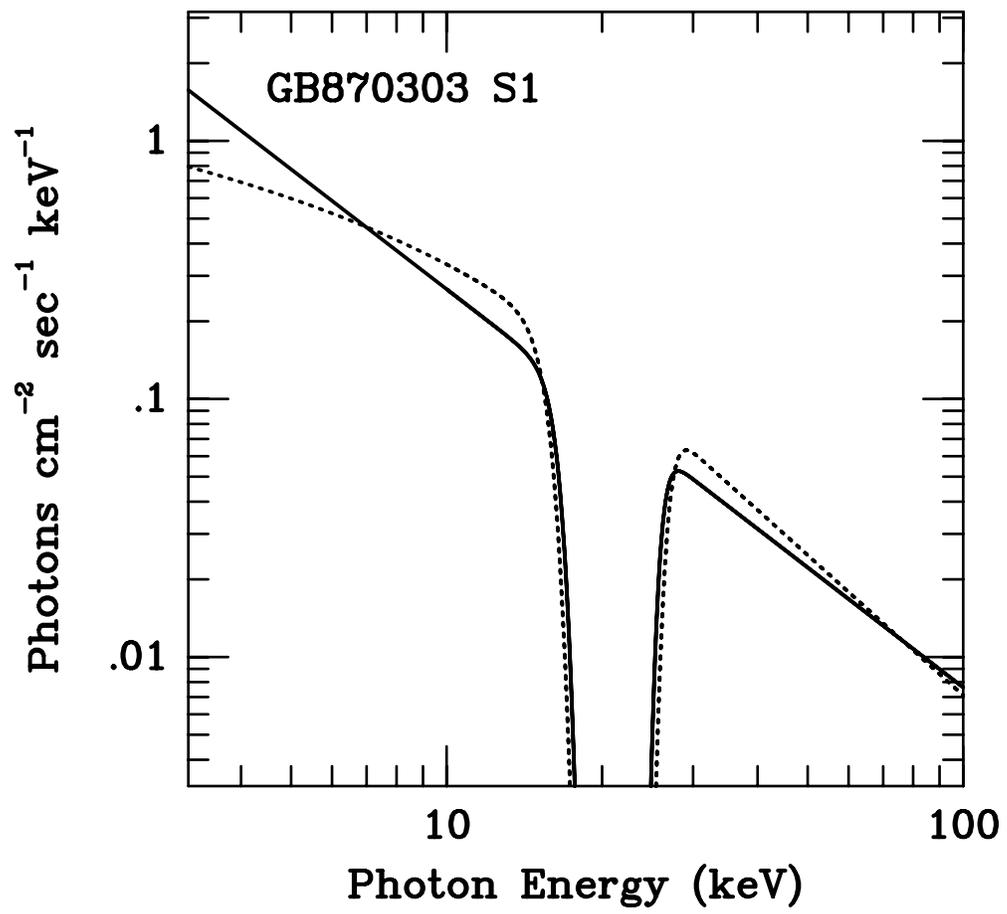
\end{document}